\documentclass[reprint,amsmath,amssymb,aps,prd,nofootinbib,showkeys]{revtex4-2}
\usepackage{amssymb}
\usepackage{lipsum}
\usepackage{xcolor}
\usepackage{graphicx}
\usepackage{amsmath}
\usepackage{mathrsfs}
\usepackage[utf8]{inputenc} 
\usepackage[T1]{fontenc} 
\usepackage{dcolumn}
\usepackage{bm}
\usepackage{orcidlink}
\usepackage{manuscript}
\usepackage{amsfonts} 
\usepackage{microtype}
\usepackage{booktabs}
\usepackage{tabularx}
\usepackage{nicefrac} 
\usepackage{xcolor}
\usepackage{url}
\usepackage{ragged2e}
\definecolor{deeppink}{RGB}{255,20,147}
\usepackage{natbib}
\setcitestyle{authoryear,round}
\usepackage{enumitem}
\usepackage{hyperref}
\hypersetup{
    colorlinks=true,
    linkcolor=deeppink,
    citecolor=deeppink,
    urlcolor=deeppink}

\renewcommand{\date}[1]{\def\@date{#1}}

\begin{document}

 \date{}

\title{\huge{Dipolar fluence distribution of statistically isotropic FERMI gamma-ray bursts}}

\author{Maria Lopes\orcidlink{0000-0001-9181-5675}}
\email{marialopes@onbr}
\affiliation{Observatório Nacional, Rua General José Cristino 77, São Cristóvão, 20921-400 Rio de Janeiro, RJ, Brasil}

\author{Armando Bernui\orcidlink{0000-0003-3034-0762}}
\email{bernui@on.br}  
\affiliation{Observatório Nacional, Rua General José Cristino 77, São Cristóvão, 20921-400 Rio de Janeiro, RJ, Brasil}

\author{Wiliam S. Hipólito-Ricaldi\orcidlink{0000-0002-1748-553X}}
\email{wiliam.ricaldi@ufes.br}  
\affiliation{Departamento de Ciências Naturais, Universidade Federal
do Espírito Santo, Rodovia BR 101 Norte, km. 60,
29932-540, São Mateus, ES, Brasil}
\affiliation{Núcleo Cosmo-UFES,  Universidade Federal do Espírito Santo,
Av. Fernando Ferrari, 540, 29075-910, Vitória, ES, Brasil}

\author{Camila Franco\orcidlink{0000-0002-6320-425X}}
\affiliation{Observatório Nacional, Rua General José Cristino 77, São Cristóvão, 20921-400 Rio de Janeiro, RJ, Brasil}

\author{Felipe Avila\orcidlink{0000-0002-0562-2541}}
\affiliation{Observatório Nacional, Rua General José Cristino 77, São Cristóvão, 20921-400 Rio de Janeiro, RJ, Brasil}

%%%%%%%%%%%%%%%%%%%%%%%%%%%%%%%%%%%%%%%%%%%%%%%%%%%%%%%%%%%%%%

\begin{abstract}
We investigated the large-angle distribution of the gamma-ray bursts (GRBs) from the updated FERMI/GBM catalog to probe the statistical isotropy of these astrophysical transient events. 
We also studied the angular distribution of the GRB fluence as a way to explore whether this radiative feature shows some preferred direction on the sky that suggest their origin. 
Our model-independent approach performed a directional analysis of the updated FERMI/GBM catalog. 
The statistical significance of our results is obtained by comparison with a large set of statistically isotropic samples of cosmic objects, with the same features of the FERMI data.
Our analyses confirm that the angular distribution of the FERMIGRB is statistically isotropic on the celestial sphere. 
Moreover, analyzing the directional distribution of the FERMIGRB fluence, that is, the median GRB fluence in a set of directions that scans the celestial sphere, we found that this astrophysical property exhibits a net dipolar structure with a directional preference for latitudes near the Galactic plane.
However, additional studies show that this directional preference is   not correlated with the Milky Way Galactic plane, which suggests that the GRB dataset, and its fluence dipolar structure, 
are extra-Galactic in origin. 
Interestingly, the analyses of the BATSE Channel 4 fluence data, that is, those GRBs from BATSE with energy $>$ 300 keV, 
reveal that its dipole direction is very well aligned with the cosmic microwave background dipole.

\end{abstract}


\maketitle
\onecolumngrid
\noindent
\begin{center}
    \vspace{-1cm}  
    \text{(Accepted 10-Dec, 2024; to appear in A$\&$A)}
\end{center}
\twocolumngrid

%%%%%%%%%%%%%%%%%%%%%%%%%%%%%%%%%%%%%%%%%%%%%%%%%%%%

\section{Introduction}\label{sec:introduction}

The hypothesis of statistical isotropy (SI)\footnote{We  use the term isotropy to refer to statistical isotropy.} is part of the cosmological principle (CP), a criterion that supports the concordance cosmological model $\Lambda$CDM. The study of the SI with diverse cosmic probes and different methodologies is important not only because it leads to verifying the universality of the CP, but   it can also provide us with new insights into the astrophysical properties of the universe 
(see, e.g., \citealt{Colin2011,Javanmardi15,Tiwari2018,Aluri2023}). For example, analyses of the cosmic microwave background data suggest 
isotropy violation (see, e.g., \citealt{deOliveira-Costa2003,Eriksen2004a,Planck2019SI,Kester2024}). 
These results motivated the search for a possible cause, studies that include, for instance,  a nontrivial topology of the universe, primordial magnetic fields, and nonstandard inflation 
(see, e.g.,~\cite{Hipolito-Ricaldi2005,Inoue2006,Pereira07,B-HR08}). On the other hand, an isotropy violation could be related to systematics, a possibility that also deserves detailed analyses~\citep{Bernui08}. 
Thus, isotropy studies of cosmological tracers observed by new astronomical surveys or simply a reanalysis of updated catalogs are essential in order to validate the CP.

The study of the angular distribution of   gamma-ray bursts (GRBs) started in the 1990s with the Burst and Transient Source Experiment (BATSE) \citep{Fishman1994}, performed with an all-sky monitor $20-1000$~keV, one of the four instruments on board the Compton Gamma-Ray Observatory (CGRO) satellite~\citep{Gehrels93,Preece2000}. 
GRBs are astrophysical phenomena associated with  stellar catastrophic events occurring in host galaxies. 
Until recently, the astronomical community believed that the burst duration, $T_{90}$,\footnote{$T_{90}$ is the time, in seconds, used to denote the duration over which a burst emits from $5\%$ to $95\%$ of its total measured counts \citep{Koshut1996}.} 
was the only parameter   dividing the set of observed GRBs into two disjoint classes~\citep{Kouveliotou93}, thus determining the astrophysical source that originates the burst by the class to which it belongs. 
At first, the GRB sample was divided into   short-time events, $T_{90} < 2\,$s, or Short-GRBs, and   long-time events, $T_{90} > 2\,$s, or 
Long-GRBs~\citep{Kouveliotou93,Hakkila2000}. 
However, in recent years more information from these intriguing events has been collected, new studies on their radiative processes have been reported, and consequently other physical features have been recognized to complement the parameter $T_{90}$ to elucidate the astrophysical sources that  give origin to GRBs~\citep{Hakkila07,Salmon2022,Mehta24,Bargiacchi24}.

Over the last few decades the angular distribution of GRBs has been extensively examined~\citep{Meegan1992,Briggs1996,Balazs1998,Bernui08,Magliocchetti2003,Cline2005, Vavrek2008,Gibelyou2012,Li2015,Ukwatta2015,Tarnopolski2015}. 
Unlike several analyses on the isotropy of GRBs, studies of the angular distribution of their radiative properties are less common. 
Nonetheless, two recent studies have examined the angular distribution of some GRB properties such as duration, fluence, and peak flux 
in the FERMI dataset: \citet{Ripa2017} analyzed a sample of 1591 GRBs from FERMI data, finding a marginal anisotropic signature in the peak flux and fluence with low statistical significance.
However, in an updated 
study,~\cite{Ripa2018} analyzed a sample of 2266 GRBs from FERMI 
and $\sim\!\!2000$ GRBs from BATSE, finding those properties to be consistent with isotropy, suggesting that possible selection effects might cancel out potential anisotropic features.

For this work we used the latest FERMI/GBM dataset to analyze the angular and fluence distributions of the GRB on the celestial sphere, applying a coordinate-free method in spherical caps \citep{Bernui07,Kester2024}. 
We note that the FERMI dataset does not provide information regarding the GRB redshift because, in general, it is not possible to observe these fast events with telescopes to perform spectroscopy of their visible counterpart. 
For this, all our analyses of the FERMI catalog deal with the angular distribution of GRB events and their measured fluence. 
The current FERMI catalog comprises 3703 GRBs from the FERMI/GBM 
instrument~\citep{Meegan09}. 
In addition, we also consider the BATSE dataset~\citep{batse1999}, which allows us to identify possible systematic effects.

Our directional analysis is optimal for detecting preferred directions across  the  sky. Using this method, one can identify possible directional preferences in the angular distribution of GRBs and in their fluence properties. The analysis of the angular distribution  is based on the two-point angular correlation function (2PACF), while the directional analysis of their fluence is determined by the median 
of the fluence of the set of GRB in a given direction. 
Both the 2PACF and the median were computed in GRB samples within a specific spherical cap centered at a defined direction in the sky. 
The statistical significance in both cases was assessed by comparing the results with a large set of simulated ensembles 
produced under the SI hypothesis, where we applied the same directional analysis approach.

The outline of this work is as follows. 
In the next section we briefly present the FERMI/GBM dataset and the samples used 
in this work. 
In Section~\ref{sec3} the method for the directional analysis of both the angular correlations and fluence of GRB is described, with details of the construction of isotropy catalogs necessary to quantify possible SI deviations. 
Next, in Section~\ref{sec4} we present our results. %and analyses. 
Finally, in Section~\ref{sec5} we present our conclusions and final remarks. Consistency tests to determine the robustness of our findings are presented in the Appendices.

%%%%%%%%%%%%%%%%%%%%%%%%%%%%%%%%%%%%%%%%%%%%%%%%%%%%%%%%

\section{The Fermi GBM Burst Catalog} \label{two}

The study of the angular distribution of the GRB events started with the BATSE catalogs released at the end of the last century~\citep{Gehrels93}. 
However, the investigation of the angular distribution of the GRB radiative properties is more recent, motivated by the importance of these studies for a better comprehension of the stellar evolution in galaxies~\citep{Gruber2014,Hakkila2000,Mehta24}.

\begin{figure}[htbp]
\centering
\includegraphics[width=\columnwidth]{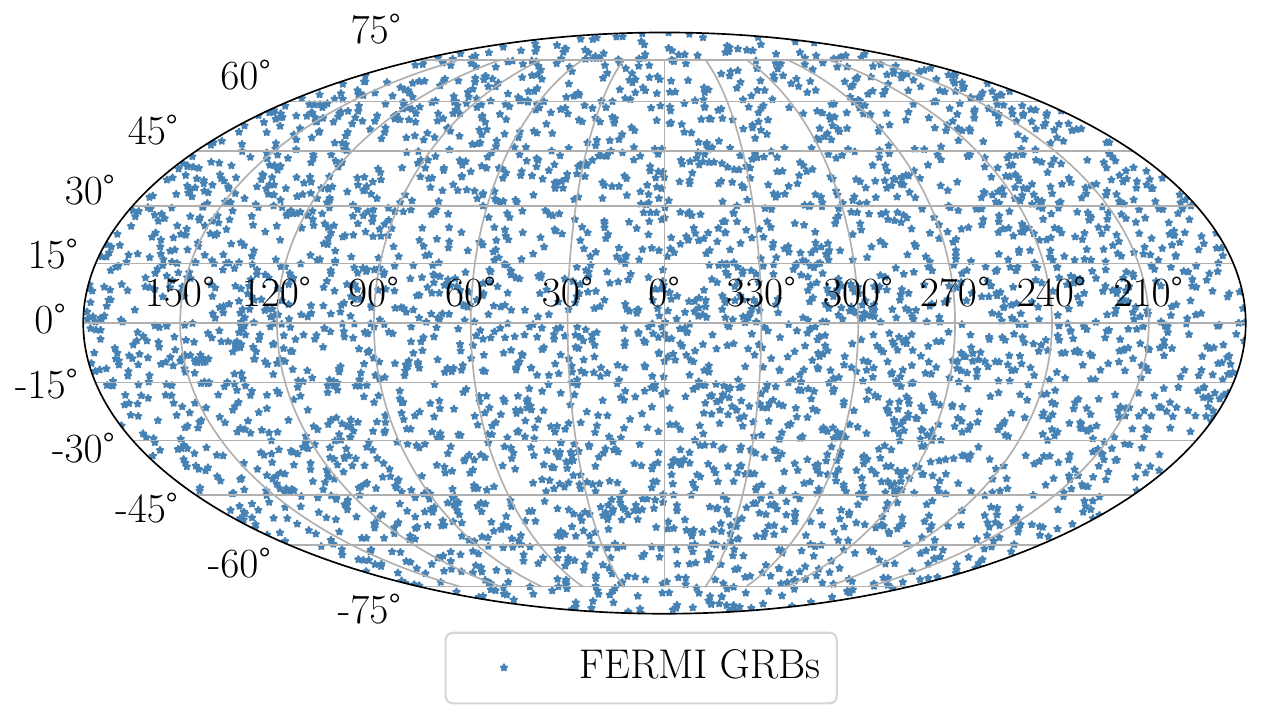}
\caption{Distribution on the celestial sphere of $3703$ GRB of the FERMIGRB 
catalog (in Mollweide projection, in Galactic coordinates).} 
\label{fig:mollweide-projection}
\end{figure}

%--------------------------------------------------------------------------------%

For our analyses we used the current GRB catalog from the FERMI Gamma-ray Burst Monitor instrument (FERMI/GBM)\footnote{\url{https://heasarc.gsfc.nasa.gov/W3Browse/fermi/fermigbrst.html}} \citep{von2020}, which is continuously being updated. 
In this study we used the version made available in February 2024, which contains 3703 GRBs detected and their fluence measured; we   refer to this sample of 3703 GRBs as the FERMIGRB catalog. 

For our studies we made use of the following quantities of the FERMI GBM Burst catalog: RA, right ascension, and  DEC, declination, both given in J2000 decimal degree units; fluence, the  flux integrated over the burst duration in erg~cm$^{-2}$ in the range 10--1000 keV energy band; and $T_{90}$.

To investigate the SI of the FERMIGRB catalog, 
we performed a directional analysis of the large-angle correlations in the celestial sphere. 
For this, we  studied three samples of GRBs: the full FERMIGRB catalog plus two  subsamples obtained by considering $T_{90}$ as the division criterion: 
GRBs with $T_{90} < 2\,$s termed short gamma-ray bursts (SGRBs), and GRBs with $T_{90} > 2\,$s termed long gamma-ray bursts (LGRBs). 
The sky distribution of the FERMIGRB sample is shown in 
Figure~\ref{fig:mollweide-projection}. 
The analyses of the subsamples of SGRBs and LGRBs are 
presented in Appendix~\ref{sigma-S-L}; in particular, their
sky distributions are shown in Figure~\ref{fig:mollweide-SLGRBs}.

In addition, we investigated the angular distribution of the fluence of the FERMIGRB sample, the fluence data of the FERMIGRB in the energy range of the BATSEGRB (i.e., 50--300~keV which we  refer to as the FERMI$_{\text{BAT}}$ sample), and 
for the sake of comparison  the fluence of the Fourth BATSE Gamma-Ray Burst Catalog\footnote{\url{https://heasarc.gsfc.nasa.gov/w3browse/all/batsegrb.html}} \citep{Paciesas1999}, which contains $1637$ GRBs. 
However, for our analysis, we selected only the events for which the fluence was recorded simultaneously in all four GRB channels. 
In the case of the BATSE data, which we  refer to as the BATSEGRB sample, we used the fluence measurements obtained in four energy channels: 
\begin{itemize}[itemsep=0pt, topsep=0pt, parsep=0pt, left=0pt, labelsep=1em]
\item
Channel 1$-$BATSEGRB, with energy range 20--50 keV; 
\item
Channel 2$-$BATSEGRB, with energy range 50--100 keV; 
\item
Channel 3$-$BATSEGRB, with energy range 100--300 keV; 
\item
Channel 4$-$BATSEGRB, with energy range $>$ 300 keV.
\end{itemize}

%%%%%%%%%%%%%%%%%%%%%%%%%%%%%%%%%%%%%%%%%%%%%%%%%%%%%%%%%%%%%%%

\section{Methodology}\label{sec3}

In this section we describe the methodology adopted for the directional analysis that measures the large-scale angular correlations of the FERMIGRB sample, and also the directional analysis of the GRB fluence from FERMI and BATSE GRB catalogs. 

%%%%%%%%%%%%%%%%%%%%%%%%%%%%%%%%%%%%%%%%%%%%%%%%%%%%%%%%%%%

\subsection{Angular distribution of the GRB: 
\texorpdfstring{$\sigma$}{sigma}-map}\label{section3.1}

Our main goal is to search for possible preferred directions in the sky distribution of the FERMIGRB, and for this we studied the intensity of their large-angle correlations. 
Our directional analysis is model-independent and uses as estimator the 2PACF 
of Landy-Szalay (LS,~\cite{LS93}; see also~\cite{deCarvalho20,Keihanen19,Avila19,Avila21,Franco24b}).

We define a spherical cap, with radius equal to $\gamma$, centered in the $J$ pixel with angular coordinates $(\theta_J, \phi_J)$ as 
\begin{equation}\label{eq:sph_cap}
\Omega_{\gamma}^J \equiv \Omega(\theta_J, \phi_J; \gamma) \subset \mathcal{S}^2 \,,
\end{equation}
where $\mathcal{S}^2$ represents the celestial sphere, and $J$ is the index of the cap in analysis, with $J = 1,...,N_{\text{caps}}$. 
We note that a hemisphere is a spherical cap 
with $\gamma = 90^{\circ}$. In each   of these caps, the 2PACF is computed using the LS estimator 
\begin{equation}\label{eq:2pacf}
\omega(\theta)^J \equiv \frac{DD(\theta) - 2DR(\theta) + RR(\theta)}{RR(\theta)}\,,
\end{equation}
with $\theta \in \langle\, 0, 2\gamma\, ]$;\footnote{The symbol $\langle \cdots,\cdots ]$ means that the interval is open by the left and closed by the right.} 
$DD(\theta)$ is the number of GRB pairs in the sample data with angular separation $\theta$, normalized by the total number of pairs; $RR(\theta)$ is a similar quantity, but for the pairs in a random simulated sample; and $DR(\theta)$ corresponds to a cross-correlation between a data object and a random object. 
Our directional analysis only uses the measured angular coordinates (i.e., the sky position, of the GRB event to calculate the angular distance between pairs), which is given by
\begin{equation}\label{eq:theta}
\theta_{ij} = \cos^{-1}[\sin(\delta_i)\sin(\delta_j) + \cos(\delta_i)\cos(\delta_j)\cos(\alpha_i - \alpha_j)] \,,
\end{equation}
where $\alpha_i$, $\alpha_j$ and $\delta_i$, $\delta_j$ are the right ascension and the declination, respectively, of the GRB $i$ and $j$.

We denote by $\omega_k^J \equiv \omega(\theta_k)^J$ the 2PACF value, in the $J$-th spherical cap, for the angular distances $\{ \theta_{ij} \}$ in the interval $\theta_{ij} \in \theta_k \equiv \langle\, (k-1)\,\beta, k\,\beta\, ]$ for $k = 1,..., N_{\textrm{\footnotesize bins}}$, 
where $N_{\textrm{\footnotesize bins}} \equiv 2\gamma / \beta$ is the number of bins and $\beta$ is the bin length. 
We adopt $\gamma = 90^{\circ}$ and $N_{\textrm{\footnotesize bins}}=35$ in our analyses. 
The number of caps, $N_{\text{\footnotesize caps}}$, depends on the pixelization parameter $N_{\text{\footnotesize side}}$; we adopt $N_{\text{\footnotesize side}} = 4$ which means $N_{\text{\footnotesize caps}} = 192$; 
we denote by $\{ n_{J} \}$, for $J = 1,...,192$, the number of GRBs in the set of $192$ hemispheres 
(i.e., $\gamma = 90^{\circ}$).

We  note that the function $\omega = \omega(\theta)^J$ measures the 
2PACF in a given direction; specifically, it encompasses the information 
regarding the angular correlations produced by the distribution of the 
cosmic objects in the $J$-th spherical cap. 
The next step is to quantify the difference between angular correlations observed in different directions. 
In order to do this we use the $\sigma$-map 
method~\citep{Bernui08,Bernui09}. 
The $\sigma$-map is a model-independent approach useful to detect preferred directions in 
data displayed on a sphere, evidencing different angular correlations intensities, analysis done through the 2PACF on spherical caps that scan the celestial sphere and then compare the observational data results with the same procedure applied to synthetic SI maps. 

Our $\sigma$-maps are celestial maps of $192$ real values, then converted into colored maps, containing information of the angular correlations intensity in each region of the celestial sphere analyzed. 
To quantify the directional variation of the correlation intensity, we defined a scalar function to associate a nonnegative real value with each spherical cap with center in $(\theta_J,\phi_J)$, that is 
\begin{equation}
\sigma_J: \Omega^J_{\gamma} \subset \mathcal{S}^2 \mapsto \mathbb{R}^+ \,,
\end{equation}
for $J = 1, ..., N_{\textrm{\small caps}}$, where our estimator $\sigma_J$ is defined by 
\begin{equation}\label{eq:sigma_j}
\sigma_J^2 \equiv \frac{1}{N_{\text{bins}}} 
\sum_{k=1}^{N_{\text{bins}}}(\omega_k^J)^2 \,,
\end{equation}
with $\sigma_J = \sigma_J(\theta_J,\phi_J)$. 
Then, the set of real positive numbers $\{ \sigma^2_J \}$, 
for $J = 1,...,N_{\text{caps}}$, is the $\sigma$-map. 
To detect any possible preferred direction or asymmetry, we perform the spherical harmonics decomposition
\begin{equation}\label{sigma1}
\sigma^2(\theta,\phi) = \sum_{\ell m} a^{\sigma}_{\ell m} Y_{\ell m}(\theta,\phi)\,,
\end{equation}
and analyze its angular power spectrum
%\text{S}
\begin{equation}\label{aps}
{\cal S}_\ell \equiv \frac{1}{2\ell + 1}
\sum_{m = -\ell}^{\ell} |a^{\sigma}_{\ell m}|^2 \,.
\end{equation}

The distribution of values of the $\sigma$-map power spectra, $\{ {\cal S}_{\ell} \}$, obtained from a similar directional analysis applied to a set of simulated SI GRB maps, helps us to quantify how frequent or rare the multipoles 
${\cal S}_{\ell}$ of the FERMIGRB $\sigma$-map are. A dipolar pattern of the map or a large value for the dipole ${\cal S}_1$, for instance, could indicate the presence of a preferred axis in the sky, that is, a possible isotropy violation in the angular distribution of the FERMIGRB. 
This makes it necessary to evaluate the statistical significance of the angular power spectra $\{ {\cal S}_{\ell} \}$ at large angles (i.e., $\ell = 1,...,5$).

%%%%%%%%%%%%%%%%%%%%%%%%%%%%%%%%%%%%%%%%%%%%%%%%%%%%%%%%%%%%%

\subsection{Angular distribution of the GRB fluence: 
\texorpdfstring{$f$}{f}-map}\label{sec:fluence-maps}

To quantify the directional variation of the fluence of the GRB sample in analysis, we consider the GRB in the $J$-th spherical cap, with center in 
$(\theta_J,\phi_J)$, and define a scalar function to associate a nonnegative 
real value, that is 
\begin{equation}
\overline{\text{F}}_{J}: \Omega^J_{\gamma} \subset \mathcal{S}^2 \mapsto 
\mathbb{R}^+ \,,
\end{equation}
for $J = 1, ..., N_{\textrm{\small caps}}$, where $\overline{\text{F}}_{J}$ is defined as the median fluence 
\begin{equation}\label{eq:median-fluence}
\overline{\text{F}}_{J} \equiv \text{median} [\{ F_i \}_J] \,,
\end{equation}
being $\{ F_i \}_J$ the set of fluences of the GRB located in the $J$-th spherical cap.

The set of $N_{\text{caps}}$ values, $\{ \overline{\text{F}}_{J} \}$, 
for $J = 1, ..., N_{\textrm{\small caps}}$, is then 
assembled together into a full-sky map (hereafter the fluence-map 
or $f$-map). Then this map is decomposed in spherical harmonics and its angular power spectrum analyzed as usual 
\begin{equation}\label{apsF}
{\cal F}_\ell \equiv \frac{1}{2\ell + 1}
\sum_{m = -\ell}^{\ell}|a^{\overline{\text{F}}}_{\ell m}|^2 \,.
\end{equation}
Finally, the evaluation of the statistical significance of the angular power spectrum $\{ {\cal F}_\ell \}$ provides us with a measure of the possible isotropy deviations, at several scales, of the angular distribution of the GRB fluence.

%%%%%%%%%%%%%%%%%%%%%%%%%%%%%%%%%%%%%%%%%%%%%%%%%%%%%%%%%%%%%%

\subsection{Simulating isotropic 
\texorpdfstring{$\sigma$}{sigma}-maps and \texorpdfstring{$f$}{f}-maps}  \label{sec:isotropic-maps}

An important part of our directional analysis is the evaluation of the multipole features of the $\sigma$-maps and $f$-maps. 
This is done by quantifying their angular power spectra and comparing them with those spectra computed from large sets of $\{ \sigma^{\text{ISO}} \}$-maps and $\{ f^{\text{ISO}} \}$-maps 
produced under the SI hypothesis and according to the following procedures.

First, we recall that in the FERMIGRB catalog the 
set of numbers of GRB in each hemisphere is given by 
$\{ n_{J} \}$, $J = 1,...,192$, corresponding to $192$ hemispheres. 
For the $\sigma$-map analysis, we first generate a random catalog with 
$100\,000$ objects uniformly distributed on the celestial sphere, that is, considering $\alpha \in [0^{\circ}, 360^{\circ}]$ and 
$\delta \in [-90^{\circ}, 90^{\circ}]$ (see, e.g.,~\cite{Franco24}). 
From this dataset we randomly select $n_{J}$ simulated cosmic objects for the $J$-th hemisphere, and  
then we obtain the set of numbers $\{ n_{J} \}$ for the $192$ hemispheres. Next, we apply our directional analysis to this ensemble 
 obtaining one $\sigma^{\text{ISO}}$-map. 

\begin{figure*}[htbp]
\begin{center}
\includegraphics[width=0.32\textwidth] {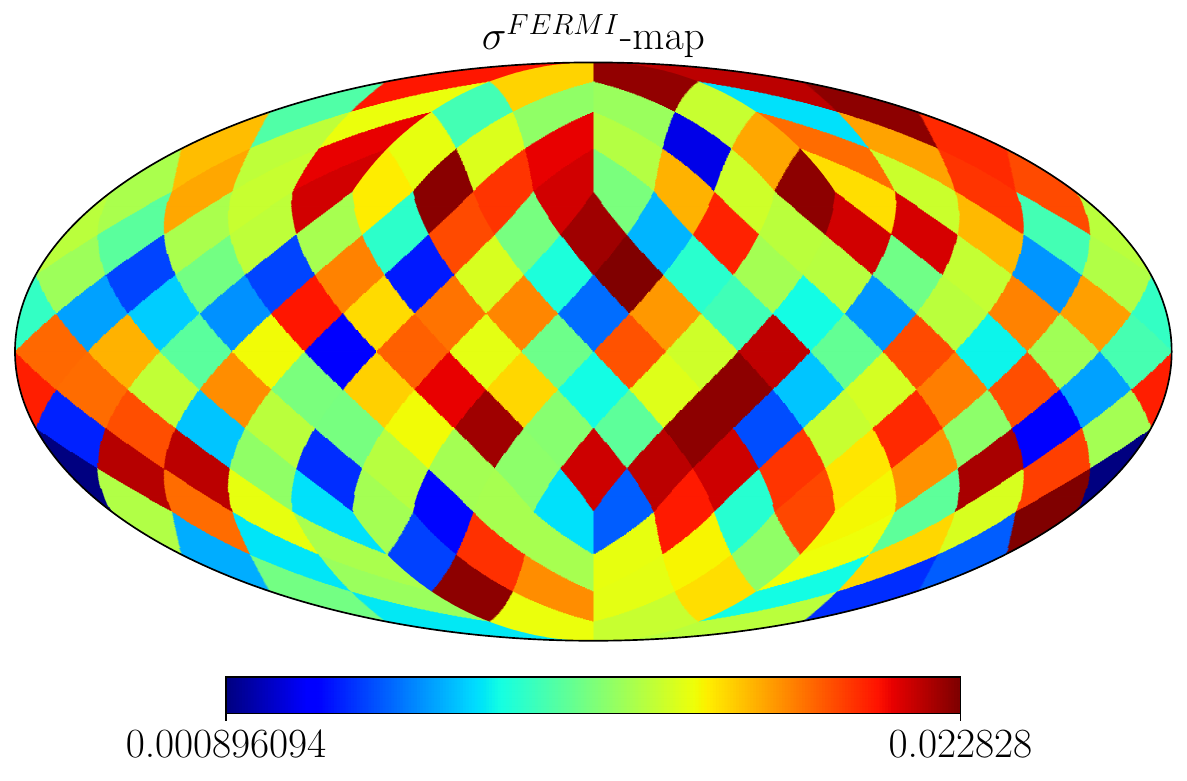}
\includegraphics[width=0.32\textwidth] {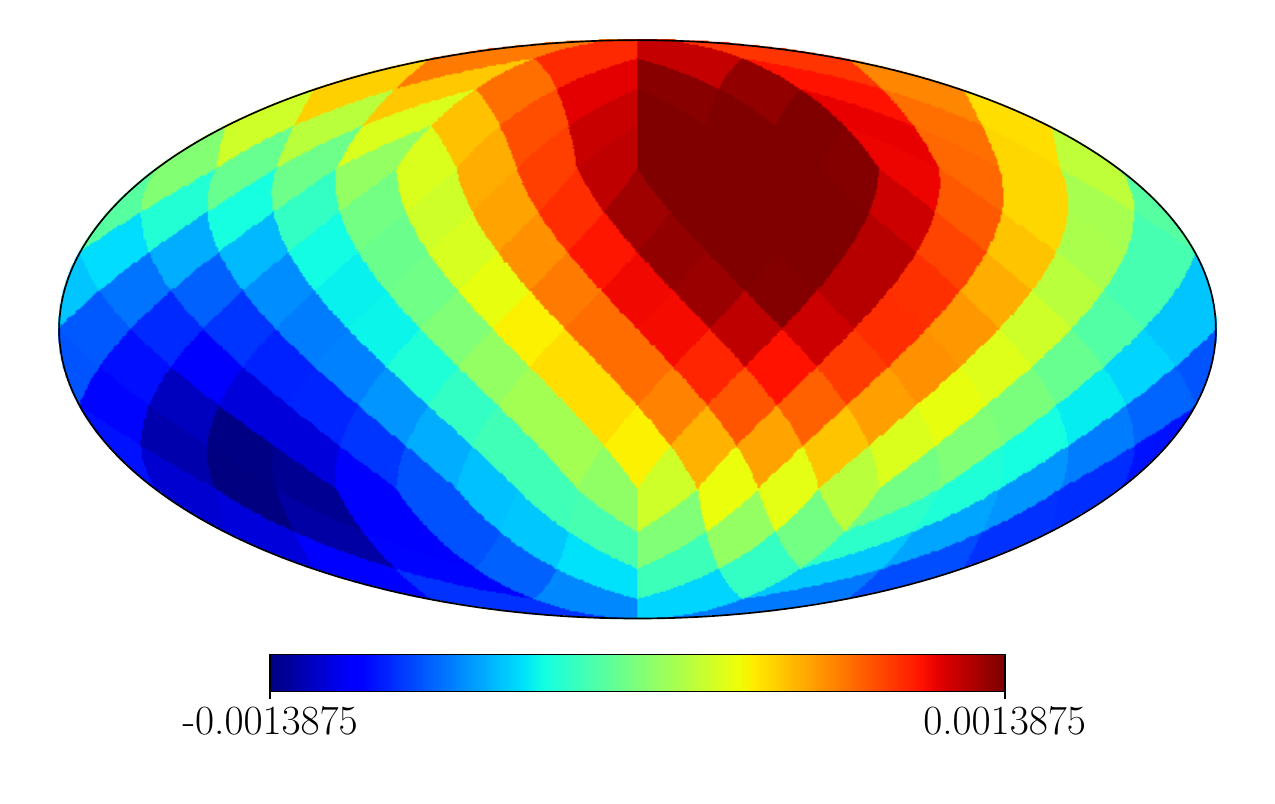} 
\includegraphics[width=0.33\textwidth]{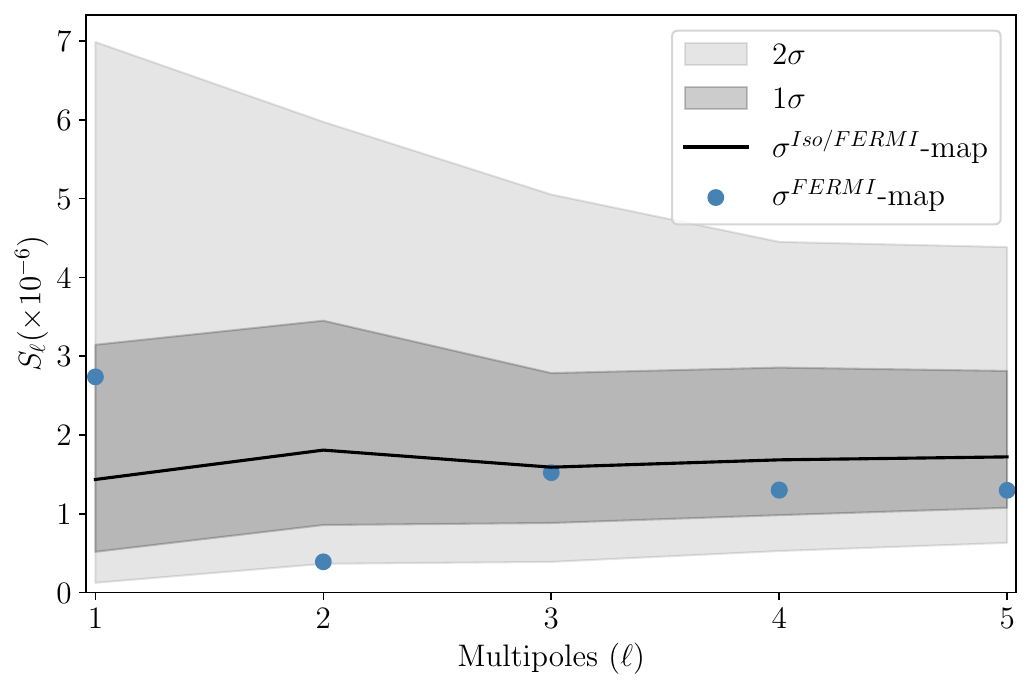} 
\end{center}
\caption{Directional analysis of the GRBs angular correlations. \textbf{Left panel:} $\sigma^\text{FERMI}$-map, with resolution $N_{\text{side}}$=4, obtained from the directional analysis of the FERMIGRB sample. 
\textbf{Middle panel:} 
Corresponding dipole component of the $\sigma^\text{FERMI}$-map.
\textbf{Right panel:} 
Power spectrum ${\cal S}_\ell^{\text{FERMI}}$, $\ell = 1,..., 5$ of the 
$\sigma^\text{FERMI}$-map shown as blue dots, and the median power spectrum of 500 isotropic $\sigma$-maps are shown on the solid line. The shaded region corresponds to 1$\sigma$ and 2$\sigma$ deviations from the isotropic case. 
} 
\label{sigma-GRB}
\end{figure*}

For the 2PACF analyses in each hemisphere we use the code TreeCorr\footnote{\url{https://rmjarvis.github.io/TreeCorr/_build/html/overview.html}.} written in the  Python language. In brief, this procedure corresponds to the analysis of one simulated SI GRB sky distribution, providing one $\sigma^{\text{ISO}}$-map. We repeat this procedure $500$ times, producing a set of $500$ $\sigma^{\text{ISO}}$-maps; 
then we compute the angular power spectra of each   of these maps. 
The  resulting power spectra are shown in the right panel of Figure~\ref{sigma-GRB}, where the median spectrum is 
represented as a continuous line, and the $1\,\sigma$, $2\,\sigma$ uncertainty regions are represented as shadow regions. A similar procedure of analysis was performed with the SGRB and LGRB subsamples in Appendix~\ref{sigma-S-L}, generating $500$ $\sigma^{\text{ISO/SGRB}}$-maps and $500$ $\sigma^{\text{ISO/LGRB}}$-maps, respectively.

To produce the set of $f^{\text{ISO}}$-maps we follow a different procedure. 
We keep the angular positions of each GRB, $(\alpha_i, \delta_i)$, 
therefore preserving the quantity of GRB in each hemisphere, $\{ n_{J} \}$. 
We then perform the isotropization of the fluence dataset $\{ F_i \}$ in two steps: 
$F_i \stackrel{\text{rand.}}{\rightarrow} 
F_i^{\text{ran}} \stackrel{\text{Gaus.}}{\rightarrow} F_i^{\text{ran+Gau}}$. 
In brief, we first randomize the original dataset obtaining $\{ F_i^{\text{ran}} \}$, 
and then each of these values is changed by a value randomly selected from a Gaussian distribution with mean $F_i^{\text{ran}}$ and standard deviation equal to its measured uncertainty 
$\sigma_{F_i^{\text{ran}}}$, 
obtaining $\{ F_i^{\text{ran+Gau}} \}$. The set of triplets $\{ (\alpha_i,\delta_i,F_i^{\text{ran+Gau}}) \}, \,i=1,...,3703$ form one simulated catalog, generated under the SI hypothesis, which,    after applying our directional analysis procedure,  produces one $f^{\text{ISO}}$-map. 
We repeat this procedure to finally obtain 500 $f^{\text{ISO}}$-maps.

%%%%%%%%%%%%%%%%%%%%%%%%%%%%%%%%%%%%%%%%%%%%%%%%%%%%%%%%%%%%

\section{Analysis and results} \label{sec4}

We applied the methodology described in the previous section to the 
FERMIGRB catalog. 
The results of the directional analysis are discussed separately: 
first we analyze the angular distribution of the GRB, and then their 
fluence distribution. 
In both cases the statistical significance was evaluated by comparison with similar analyses applied to isotropic synthetic Monte Carlo catalogs ($\sigma^{\text{ISO}}$-maps and $f^{\text{ISO}}$-maps) constructed as described above (see Sect. \ref{sec:isotropic-maps}). These analyses allowed us to quantify the significance level of our findings and to evaluate whether the distribution of FERMIGRB and their respective fluences are isotropic.

%%%%%%%%%%%%%%%%%%%%%%%%%%%%%%%%%%%%%%%%%%%%%%%%%%%%%%%%%%%%%%%

\subsection{Gamma-ray bursts distribution on the sky}\label{GRB-sky-dist}

We investigated the large-angle distribution of FERMIGRB on the sky. 
This was done through the analysis of the intensity of the angular correlations in several directions along the celestial sphere, as explained in Sect.~\ref{section3.1}. 

We first calculated the $\sigma$-map for the full GRB sample, $\sigma^{\text{FERMI}}$-map. 
Our result is displayed in the left panel of Figure~\ref{sigma-GRB}, 
where it can be seen that there is no excess or deficit of correlation around any particular direction. 
In other words, correlation intensities are isotropically distributed around the sky. 
In the middle panel we show the $\sigma^{\text{FERMI}}$-map dipolar component. Next, we quantified the significance of the large-angle multipoles, through a comparison with a set of $500$ SI 
$\sigma$-maps. To do this, we generated a set of $500$ simulated SI GRB catalogs and applied the same directional analysis methodology 
to each one, obtaining a set of $500$ 
$\sigma^{\text{ISO}}$-maps. 
Then, we calculated their angular power spectra ${\cal S}_\ell$, for $\ell=1,...,5$. 
Finally, we compared the spectral angular signatures of the $\sigma^{\text{FERMI}}$-map (i.e., ${\cal S}_\ell^{\text{FERMI}}$) with these SI spectra. 
The result is displayed in the right panel of Figure~\ref{sigma-GRB}, 
where ${\cal S}_\ell^{\text{FERMI}}$ is represented by blue dots, the black line 
represents the median value of the set $\{ {\cal S}_\ell^{\text{ISO}} \}$; 
the 1$\sigma$  and  2$\sigma$ regions are shown   shaded in gray. 
We observe that the  $\sigma^{\text{FERMI}}$-map is compatible with the SI hypothesis, within the  $1\,\sigma$ confidence level, at all the large-angles analyzed. 
Moreover, as a consistency test, we analyzed the angular distribution for the SGRB and LGRB subsamples, with the discussion presented in the Appendix~\ref{sigma-S-L}.

%%%%%%%%%%%%%%%%%%%%%%%%%%%%%%%%%%%%%%%%%%%%%%%%%%%%%%%%%%%%%%

\subsection{Gamma-ray bursts fluence distribution on the sky}

\begin{figure*}[ht]
\begin{center}
\includegraphics[width=0.32\textwidth] {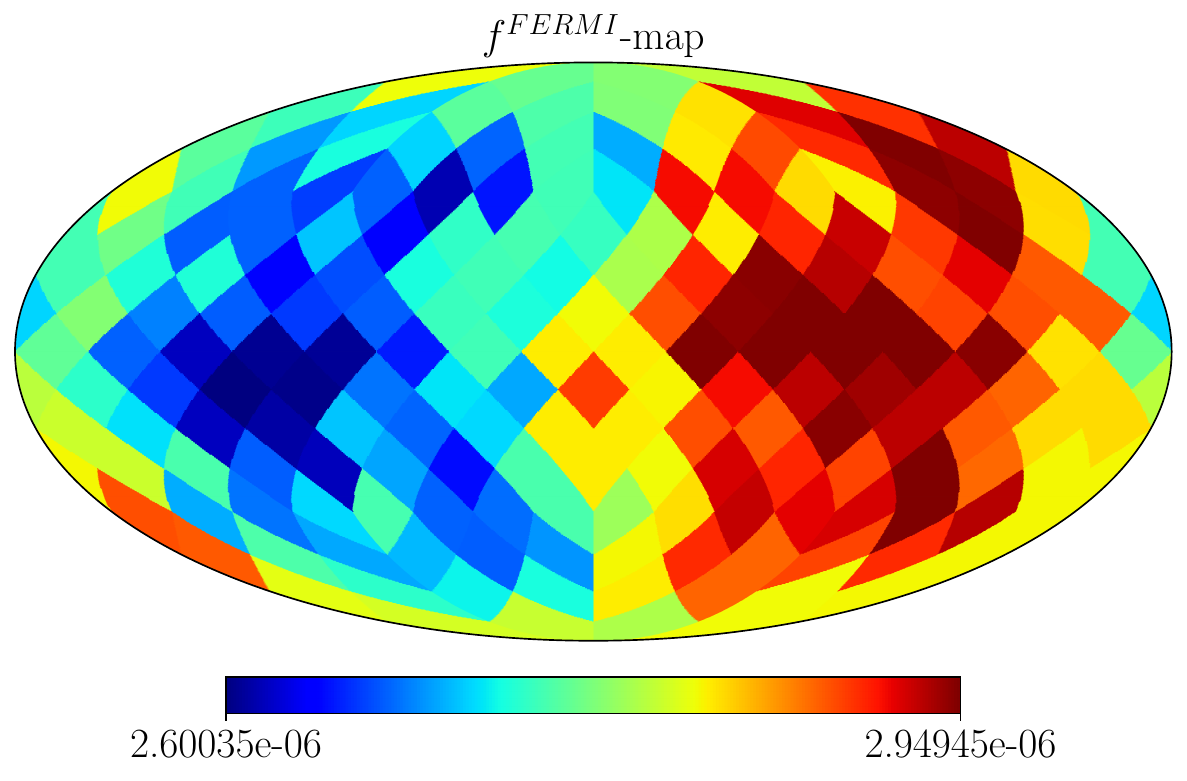}
\includegraphics[width=0.32\textwidth] {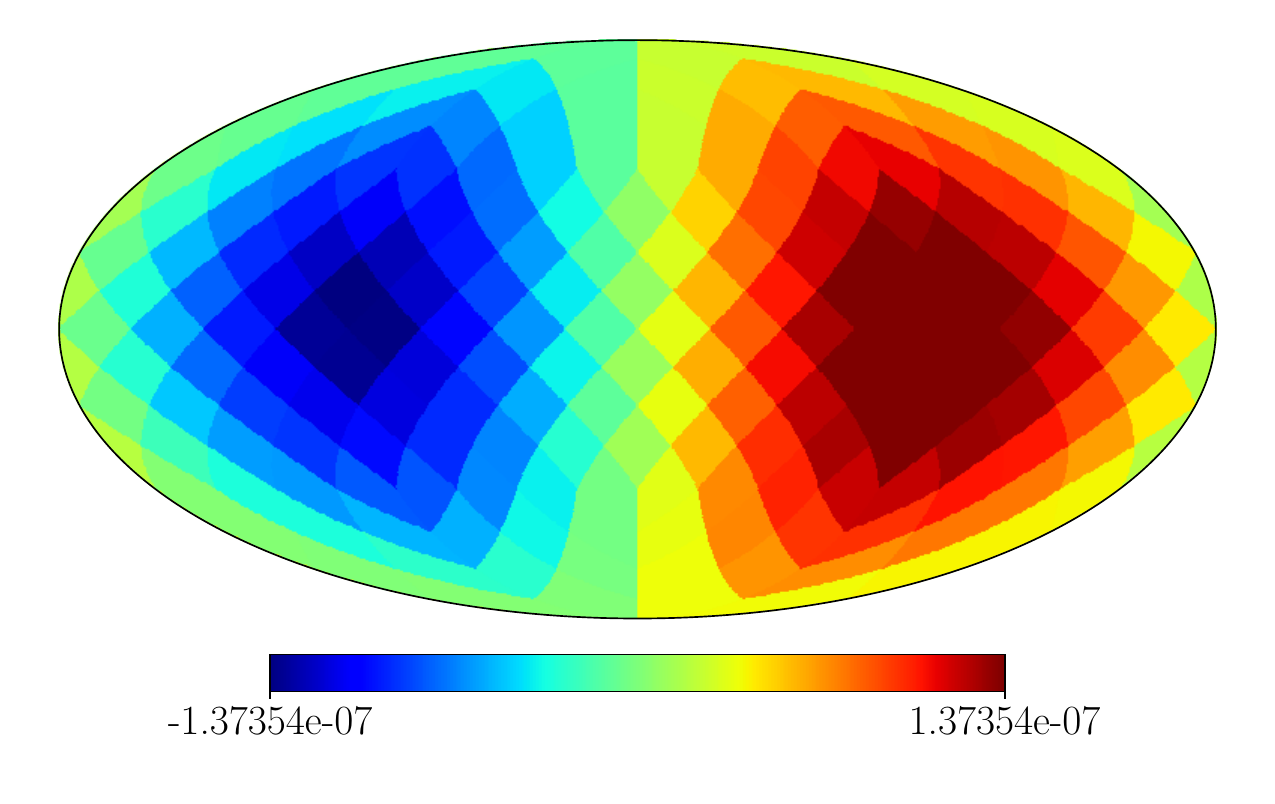} \includegraphics[width=0.33\textwidth] {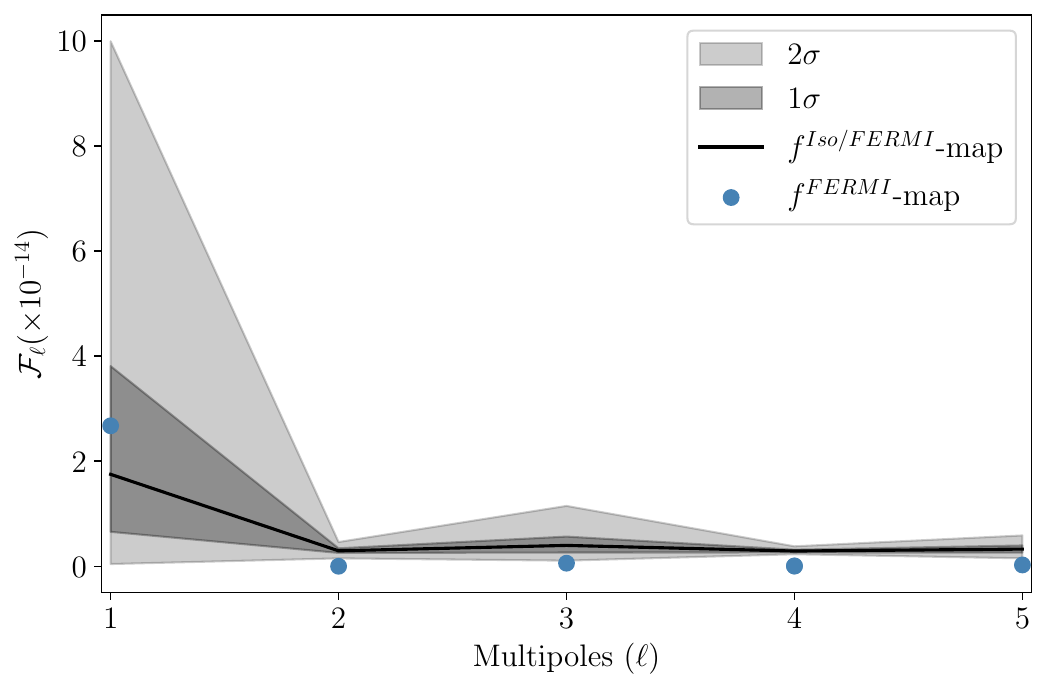}  
 \end{center}
\caption{{Directional analysis of the GRBs fluence. \bf Left panel:} $f^{\text{FERMI}}$-map 
for the FERMIGRB catalog. 
{\bf Middle panel:}
Dipole component of the $f^{\text{FERMI}}$-map. 
{\bf Right panel:} Power spectrum of the 
$f^{\text{FERMI}}$-map. 
The fluence maps are in units of erg/cm$^{2}$, in the \text{blue}{Galactic} coordinate system, 
and with resolution $N_{\text{side}} = 4$.} 
\label{figure:f_map_fermi}
\end{figure*}

\begin{figure*}[ht]
\begin{center}
\includegraphics[width=0.32\textwidth]{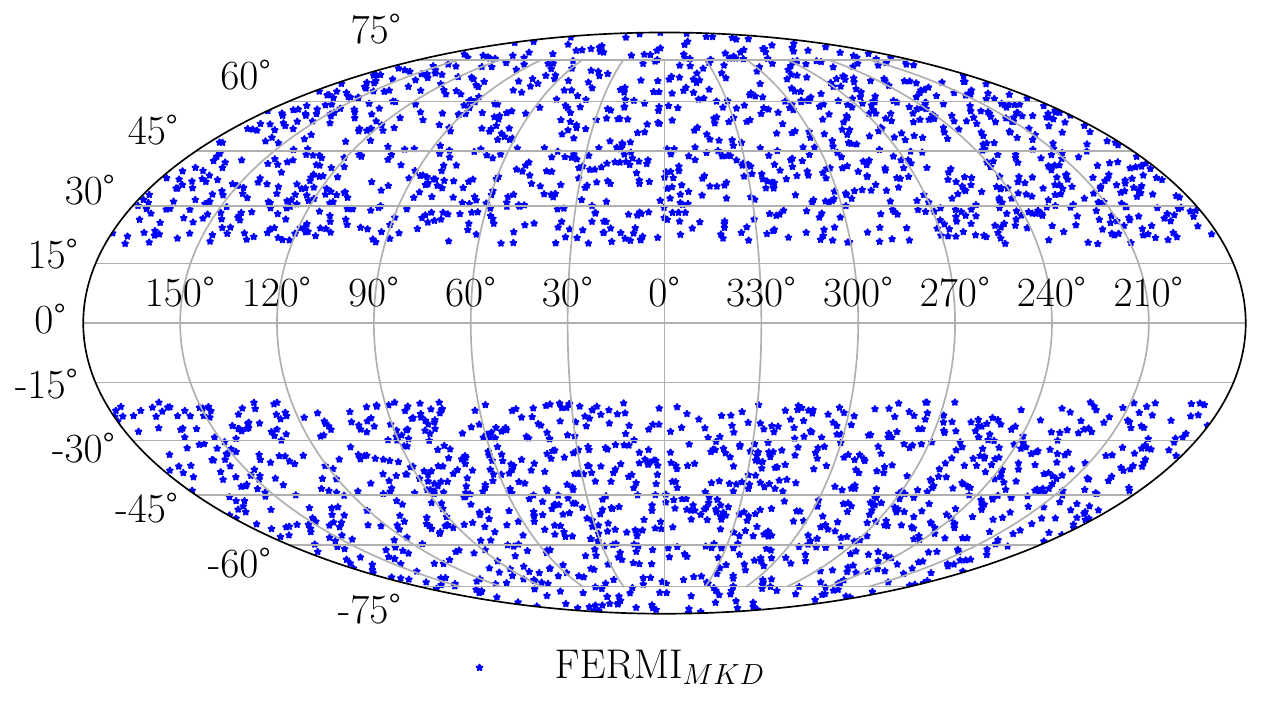}
\includegraphics[width=0.32\textwidth] {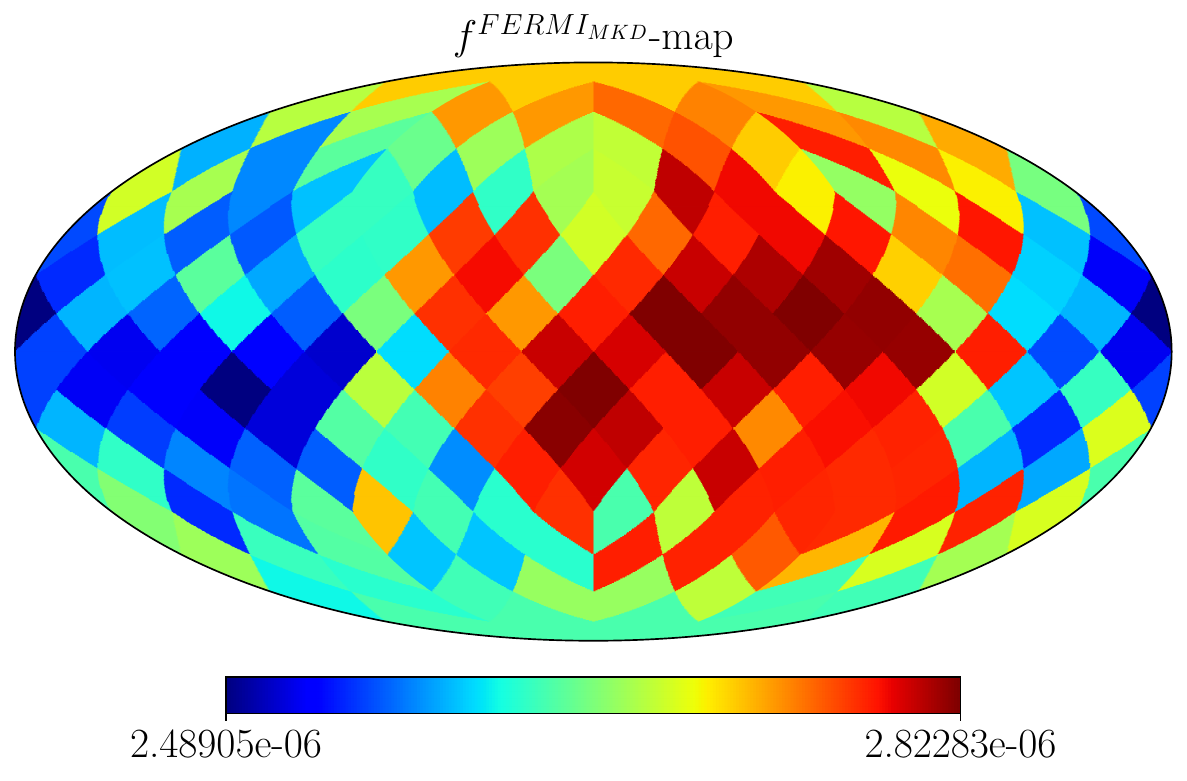}
\includegraphics[width=0.33\textwidth] {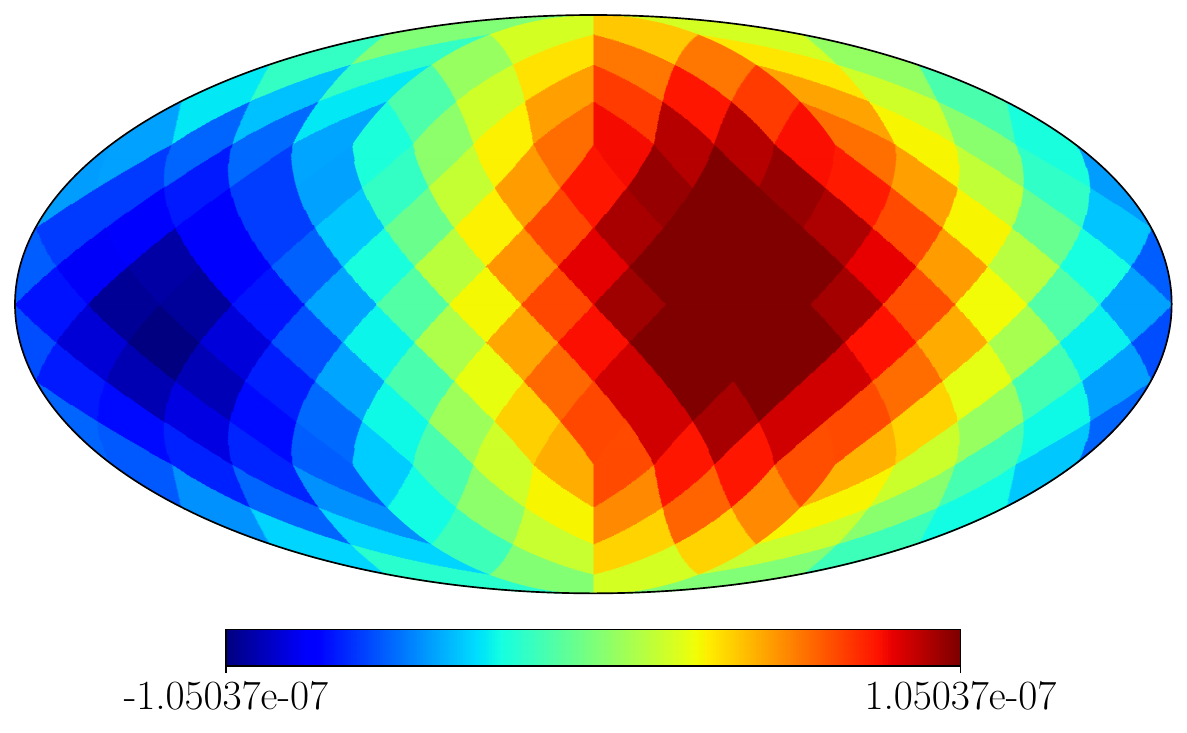}
 \end{center}
\caption{Directional analysis of the GRBs fluence: masked case. \textbf{Left panel:} 
Distribution of GRB on the celestial sphere of the FERMIGRB masked sample, showing $2428$ GRBs, where we implemented a cut of the zone of avoidance removing GRBs with $|b| \le 20^{\circ}$ (in Galactic coordinates). 
\textbf{Middle and Right panels:} 
 $f^{\text{FERMI}_{\text{MKD}}}$-map with resolution $N_{\text{side}} = 4$ (middle panel) and its dipole component (right panel) for the  FERMIGRB-Masked sample. 
All the  maps   are in units of erg/cm$^{2}$ and in the Galactic coordinate system.}
\label{f-masked-maps}
\end{figure*}

To investigate the angular distribution of the FERMIGRB fluence, we used the $f$-maps described in Sect.~\ref{sec:fluence-maps}. 
The $f$-maps allow for a directional investigation of the fluence angular distribution  across the sky, and its power spectrum, ${\cal F}_\ell$, helps us  better quantify the statistical significance of our findings. 
Figure~\ref{figure:f_map_fermi} presents the $f^\text{FERMI}$-map for the FERMIGRB catalog  (left panel) and its dipolar component (middle panel). 
The power spectrum, ${\cal F}^{\text{FERMI}}_\ell$,  shown in the right panel (blue dots) for  $\ell = 1,...,5$, indicates that the $f^{\text{FERMI}}$-map is highly dominated by its dipolar component, 
and unlike the $\sigma^{\text{FERMI}}$-map, it shows a clear dipole structure, as seen in  Figure~\ref{figure:f_map_fermi} (left panel).  
To evaluate the statistical significance of the multipole components of the 
$f^{\text{FERMI}}$-map, 
we compare them  with SI $f^{\text{ISO}}$-maps. 
To this end, we computed the angular power spectra of a set of $500$ simulated SI $f^{\text{ISO}}$-maps, whose median is shown in the right panel of 
Figure~\ref{figure:f_map_fermi} (black line), alongside the 1$\sigma$ and 2$\sigma$ regions in gray. 
We   note %in the right panel of Figure~\ref{f-maps} 
that, although the dipole behavior is remarkable, it is not anomalous compared to the isotropic dataset; the analysis of the angular power spectra shows that the $f^{\text{FERMI}}$-map is essentially a dipole. 

The dipole component from the FERMIGRB catalog indicates a net preference for the Galactic plane region, at approximately 
$(l,b) \simeq (268^\circ,-3^\circ) \pm (11.2^\circ,11.2^\circ)$. 
To investigate a possible bias in our results, we studied the correlation of our outcomes with GRBs located in the Galactic plane. 
Thus, we removed the data located in the Galactic region and repeated the analyses. 
This involved applying a cut to the region of avoidance defined as
$|b| \le 20^\circ$ in Galactic coordinates, which excluded $33 \%$ of the sky. 
The left panel in Figure~\ref{f-masked-maps} shows the distribution of GRB after applying this Galactic cut. After the cut, we computed the new $f^{\text{FERMI,MKD}}$-map, and the result is presented in the middle panel of  Figure~\ref{f-masked-maps};  its dipolar component is also shown, in the  right panel. 
The angular features and the dipolar pattern in the fluence distribution, previously found by analyzing the full-sky data, remain. This result indicates that there is no appreciable correlation between the GRB fluence localized in the Galactic plane and the dipole direction found in $f^{\text{FERMI}}$-map.

So far, we have presented in this section the analysis of the fluence of the 
FERMIGRB sample considering 
$\gamma = 90^\circ$. 
In Appendix~\ref{cap-size}, we present the  robustness tests for the directional analysis, this time considering different sizes of the spherical cap, namely $\gamma = 45^\circ, 60^\circ, 65^\circ, 70^\circ$. As shown  in Appendix~\ref{cap-size}, the dipolar pattern is consistent in all cases (see Table~\ref{table:cap-size}). 

Additionally, we investigated  the possible instrumental, systematic, or data processing effects that could bias our fluence results. 
To do this, we conducted two types of additional analyses. 
First, we evaluated possible correlations between our findings and the number of GRBs in each hemisphere using the linear Pearson correlation coefficient. 
The results, presented in   Appendix~\ref{pearson}, indicate no significant correlation. 
Second, we applied a similar directional analysis to the BATSE dataset. 
The BATSE experiment operated in four channels, with the energy intervals described in Section~\ref{two}. 
For comparison with the BATSE dataset, we constructed a subsample of the FERMIGRB catalog in an  energy range similar to that of the  BATSE data. 
The results are very similar to those displayed in Figure~\ref{figure:f_map_fermi} for the full FERMIGRB sample (see Appendix~\ref{fluence-fermi-batse}), confirming the robustness of our findings. 

%%%%%%%%%%%%%%%%%%%%%%%%%%%%%%%%%%%%%%%%%%%%%%%%%%%%%%%%%%%%%%%%%%%%%%%%%%%%%%%%%%

\section{Conclusions} \label{sec5}

An important issue of the standard cosmological model is to establish the  large-scale properties of the universe, that is, to know how matter and radiation are distributed at large scales in the observed universe~\citep{Tarnopolski2015,Dainotti22,Sorrenti24,Lopes24,Novaes2024,
Sorrenti2024c}. 
In this regard, the updated catalogs of GRBs are a valuable source of astrophysical information to inquire if such energetic events are in fact universal astrophysical processes that happen everywhere and at any epoch of the universe. 
We  investigated the large angular distribution on the celestial sphere of the latest FERMIGRB catalog, and of their fluence property, applying a model-independent approach. 
Although the first problem has been extensively studied, the second has received less attention. 

In summary, we found a highly isotropic distribution of GRB, both in the FERMIGRB catalog and in the  SGRB and LGRB subsamples (see Appendix~\ref{sigma-S-L}). 
This indicates that, on average, the GRB observed by FERMI and their host galaxies are not anomalously clustered. This reinforces their extra-Galactic origin at cosmological scales, and supports the idea that these universal astrophysical processes occur everywhere and at any epoch in the universe.

\begin{figure}[htbp]
\begin{center}
\includegraphics[width=\columnwidth] {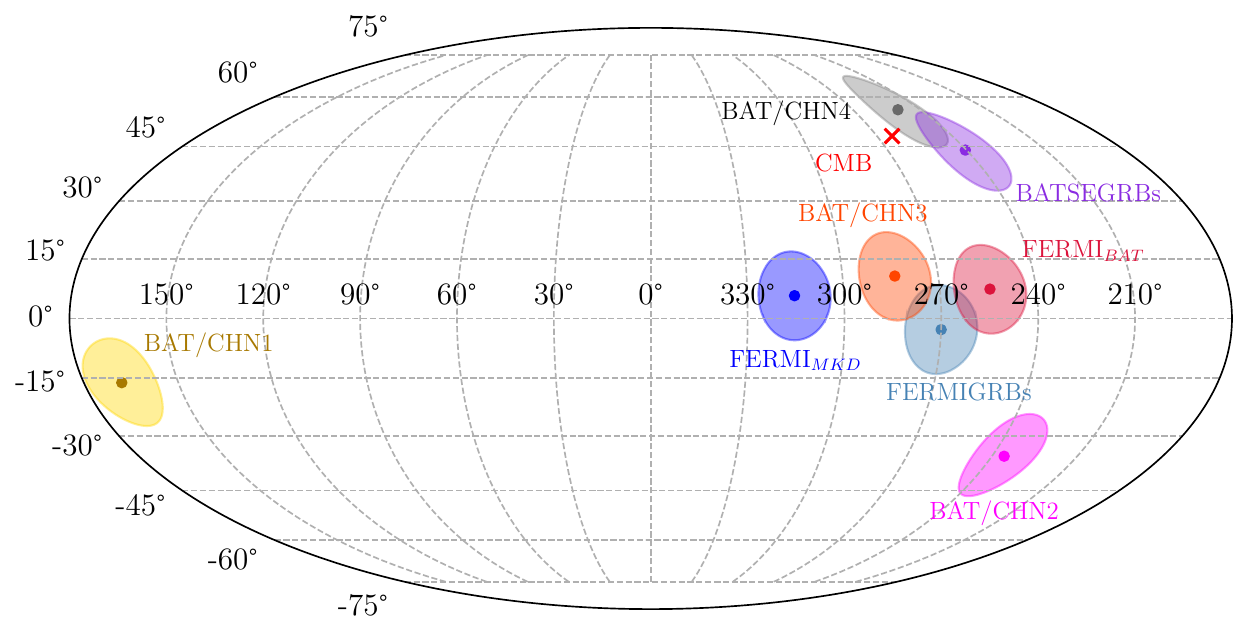}
\end{center}
\caption{Mollweide projection, in Galactic coordinates, of the dipole directions from $f$-map analyses, as displayed in Table~\ref{table:fluence-results}. 
The shaded regions corresponds to $1 \sigma$ uncertainties. 
The labels identify the samples in analysis; 
the abbreviations are as follows: 
BAT for BATSE data, MKD for masked map, and 
the 
CHN for channel in analysis.
} 
\label{fig:maps-mollw-results}
\end{figure}

Moreover, the directional analysis of the fluence data shows a net dipolar pattern 
for the FERMIGRB, FERMI$_{\text{BAT}}$, and BATSE Channel 4 catalogs, 
as revealed by their corresponding angular power spectra (see the right panels in Figures~\ref{figure:f_map_fermi} 
and~\ref{f-BATSE-maps}, 
and the analyses in Appendix~\ref{fluence-fermi-batse}), where they appear with low statistical significance, 
$\sim \!1 \sigma$, suggestive of SI. 
The dipole behavior observed in the $f$-maps of the analyzed catalogs, with their dipole direction close to the Galactic plane for the FERMIGRB and FERMI$_{\text{BAT}}$ data, motivates further consistency analyses to investigate possible systematics. 
Thus, we studied the possible correlation with the data located around the Milky Way Galactic plane or a possible influence of the number of GRB observed in each sky patch. 
Nevertheless, as shown in  Figure~\ref{f-masked-maps} and Appendix~\ref{pearson}, 
we did not find a substantial correlation for attributing to these effects the cause of the dipolar pattern. The positive dipole directions in the cases studied in this work are summarized in Figure~\ref{fig:maps-mollw-results}. Regarding the direction of the dipolar component present in the $f$-maps, we note that it roughly resembles the direction of the cosmic microwave background dipole. 
However, since the FERMI catalog does not have redshift information, it was not possible to perform a complete analysis to determine whether such a feature is a coincidence or whether it is indeed due to our motion relative to the cosmic microwave background frame.

Intriguingly, the $f$-map from BATSE Channel 4 data shows a net dipolar pattern 
with its dipole direction aligned with the direction of the cosmic microwave background dipole ~\citep{Javanmardi15,Singal19,Luongo22,Krishnan22,Zhai22,Mittal24}, 
as shown in Table~\ref{table:fluence-results} and in Appendix~\ref{fluence-fermi-batse}, although its statistical significance is not high (see Figure~\ref{f-BATSE-maps}).

%%%%%%%%%%%%%%%%%%%%%%%%%%%%%%%%%%%%%%%%%%%%%%%%%%%%%%%%%%%%%%%%%%%%%%%%%%%%%%%%%%

\begin{acknowledgments}
ML and CF acknowledge to CAPES, and AB acknowledges to CNPq, 
for their corresponding fellowships. 
WSHR thanks to FAPES (PRONEM No 503/2020) for the financial support under which this work was carried out. 
FA thanks to CNPq and FAPERJ, Processo SEI 260003/014913/2023, 
for the financial support.\end{acknowledgments}

%%%%%%%%%%%%%%%%%%%%%%%%%%%%%%%%%%%%%%%%%%%%%%%%%%%%%%%%%%%%%

\renewcommand{\bibsection}{\section*{References}}
\bibliographystyle{mnras}
\bibliography{manuscript}

%%%%%%%%%%%%%%%%%%%%%%%%%%%%%%%%%%%%%%%%%%%%%%%%%%%%%%%%%%%%

\begin{appendix}\label{appendixtotal}

%%%%%%%%%%%%%%%%%%%%%%%%%%%%%%%%%%%%%%%%%%%%%%%%%%%%%%%%%%%%%%%%%%%%%%%%%%%%%%%%%%

\section{Consistency test: The SI of the  FERMI LGRB and SGRB samples}\label{sigma-S-L}

\renewcommand{\thefigure}{A.\arabic{figure}}
\renewcommand{\tablename}{Table} 
\renewcommand{\thetable}{A.\arabic{table}}
\setcounter{table}{0}
\setcounter{figure}{0} 

In this Appendix we study the angular distribution of the Short-GRB (SGRB) and Long-GRB (LGRB) samples following the same directional analysis approach, applied to the full FERMIGRB sample (with $3703$ GRB) in Section~\ref{GRB-sky-dist}. 

\begin{figure}[ht]
\centering
\includegraphics[width=0.85\columnwidth]{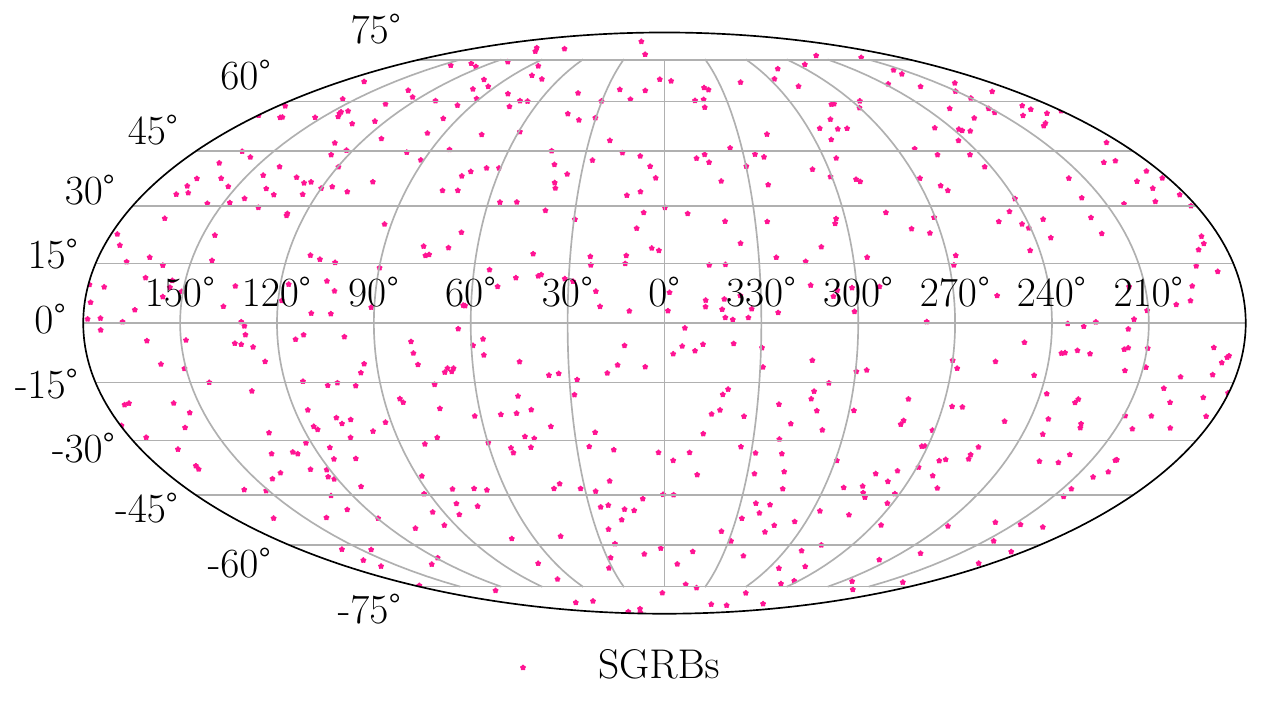} 

\includegraphics[width=0.85\columnwidth]{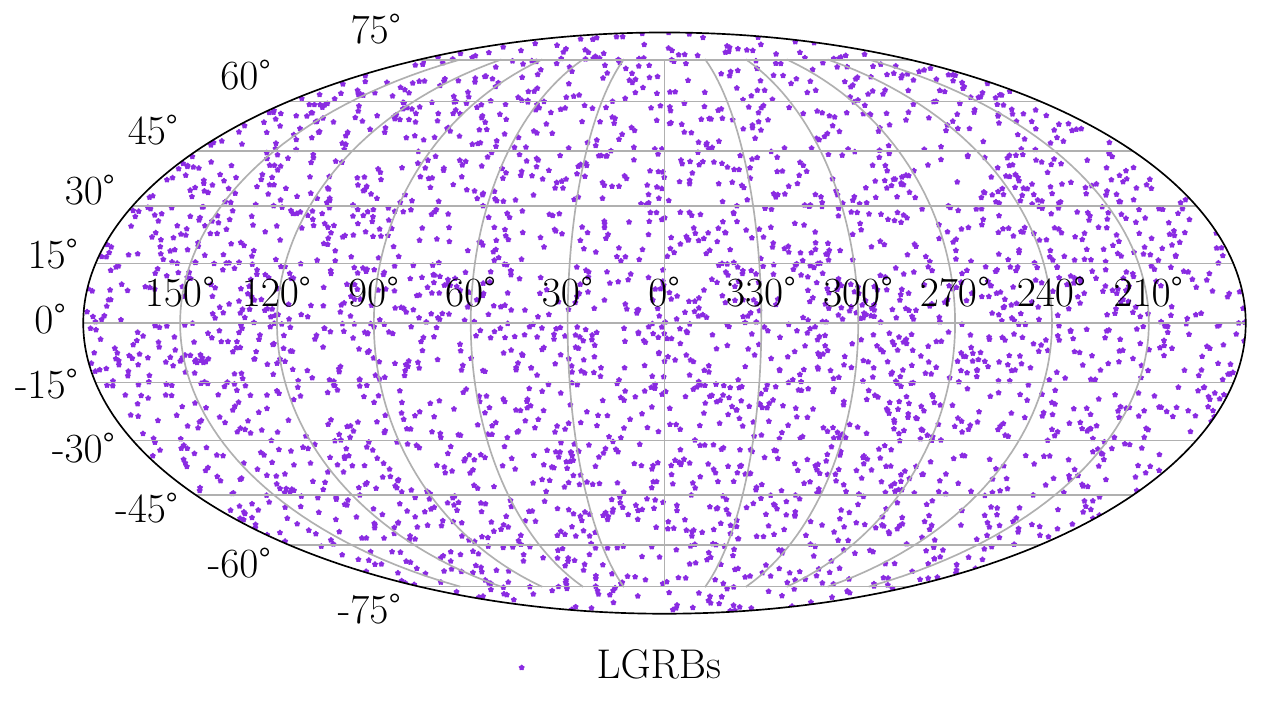}
\caption{Angular distribution in Galactic coordinates, in Mollweide projection, 
of the subsamples of $614$ SGRB (upper panel) and $3089$ LGRB (bottom panel) 
from the FERMI catalog.
} 
\label{fig:mollweide-SLGRBs}
\end{figure}

Regarding the angular distribution of the SGRB and LGRB samples, 
the literature reports analyses of different catalogs using a variety of 
methodologies (see, e.g.,~\citet{Balazs1998,Magliocchetti2003,Bernui08,Ukwatta2015, Ripa2017,Ripa2018,Tarnopolski2015}). 
For this, we find interesting to test the hypothesis of SI for each of these subsamples from the updated FERMIGRB catalog. 
In this case, the number of GRB in the SGRB and LGRB samples is $614$ and $3089$, respectively.  
The sky distribution of these samples can be seen in the Mollweide projection presented in Figure~\ref{fig:mollweide-SLGRBs}.

\begin{figure*}[htbp]
\begin{center}
\includegraphics[width=0.35\textwidth] {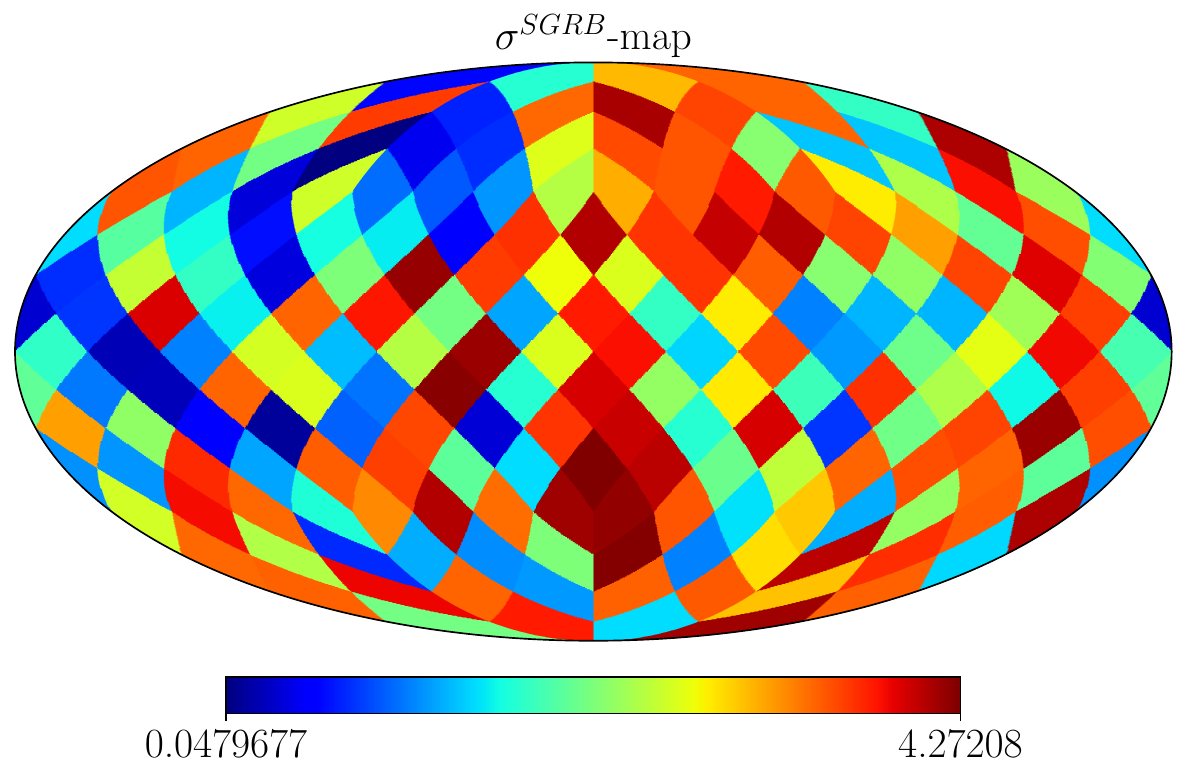}\includegraphics[width=0.35\textwidth] {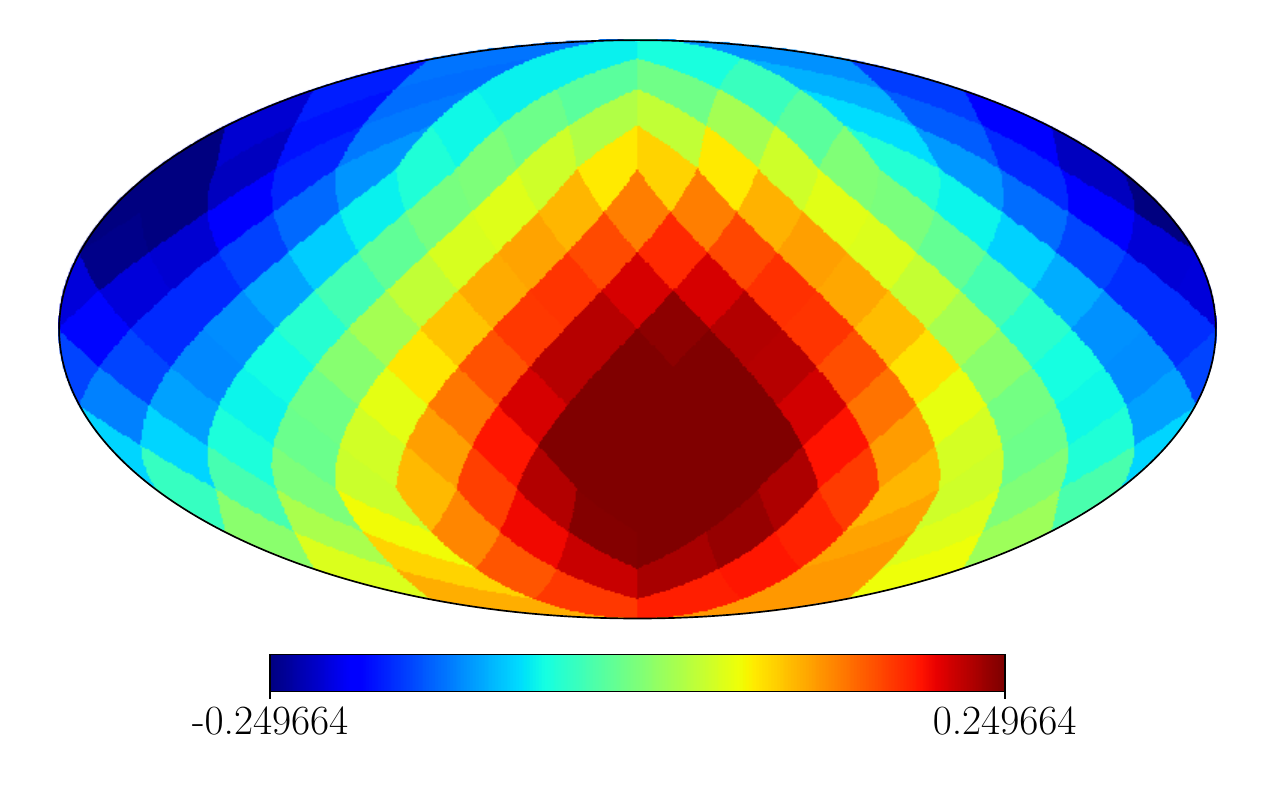}\includegraphics[width=0.35\textwidth] {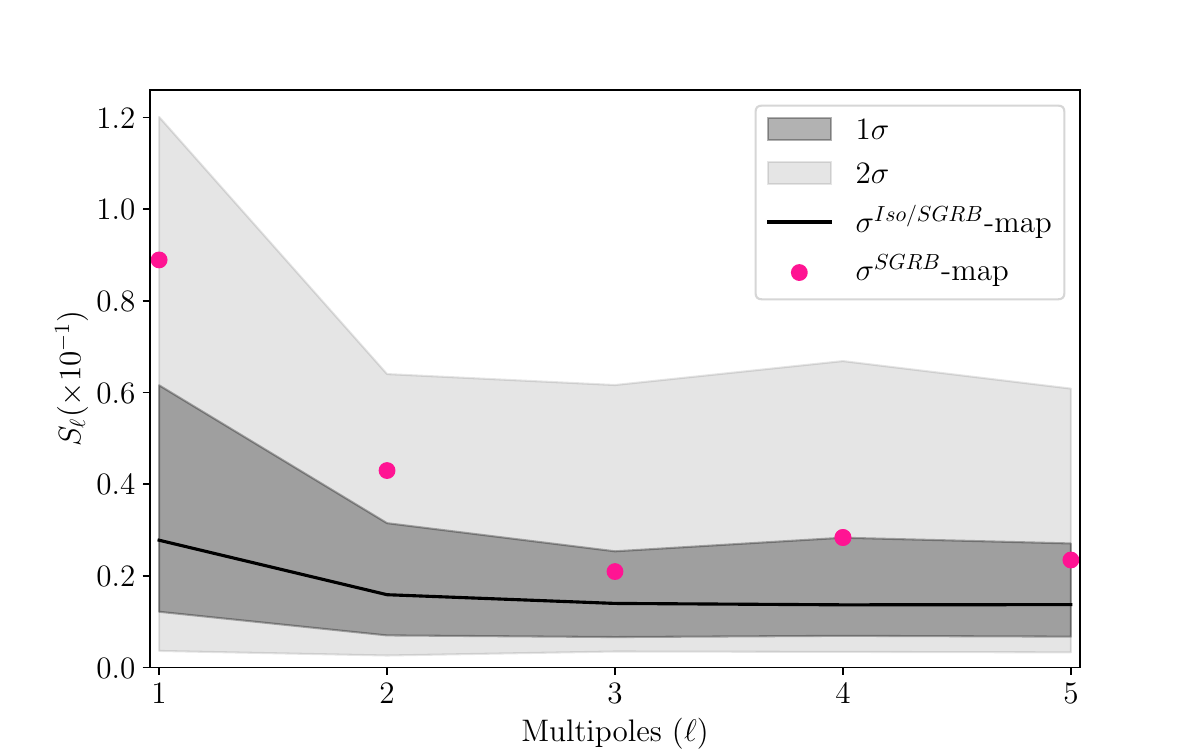} 

\includegraphics[width=0.35\textwidth] {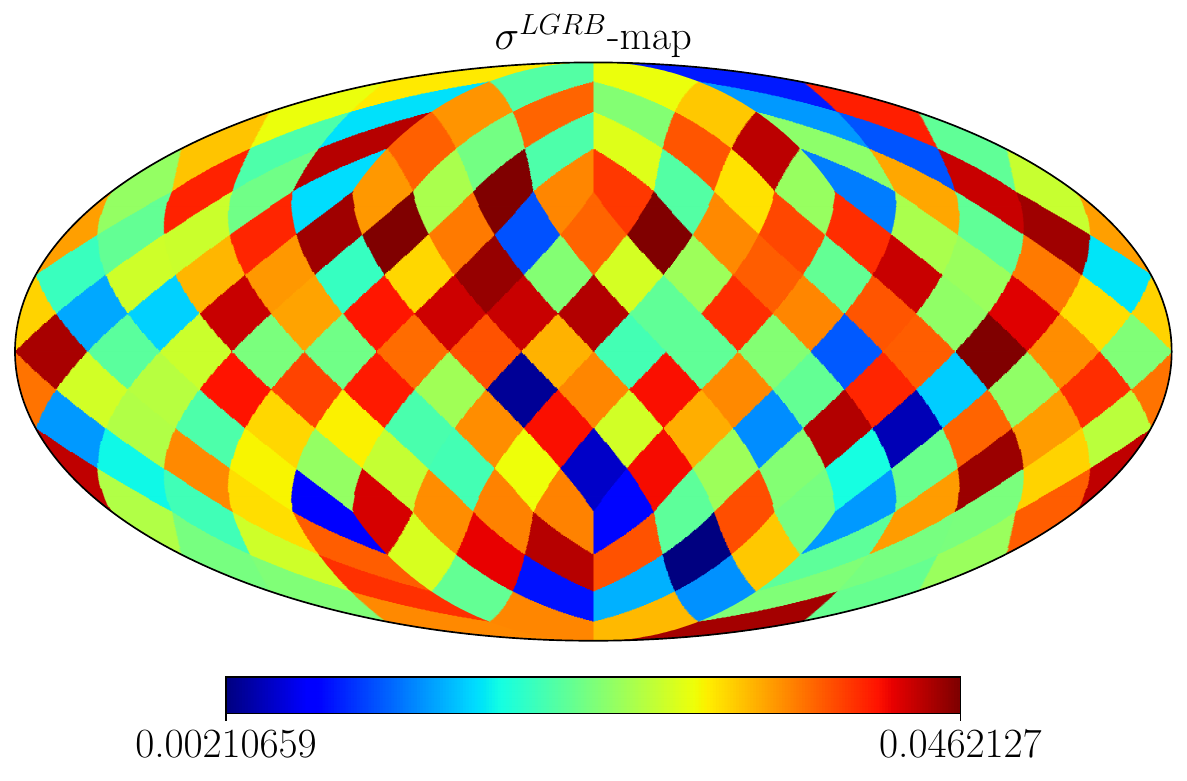}\includegraphics[width=0.35\textwidth] {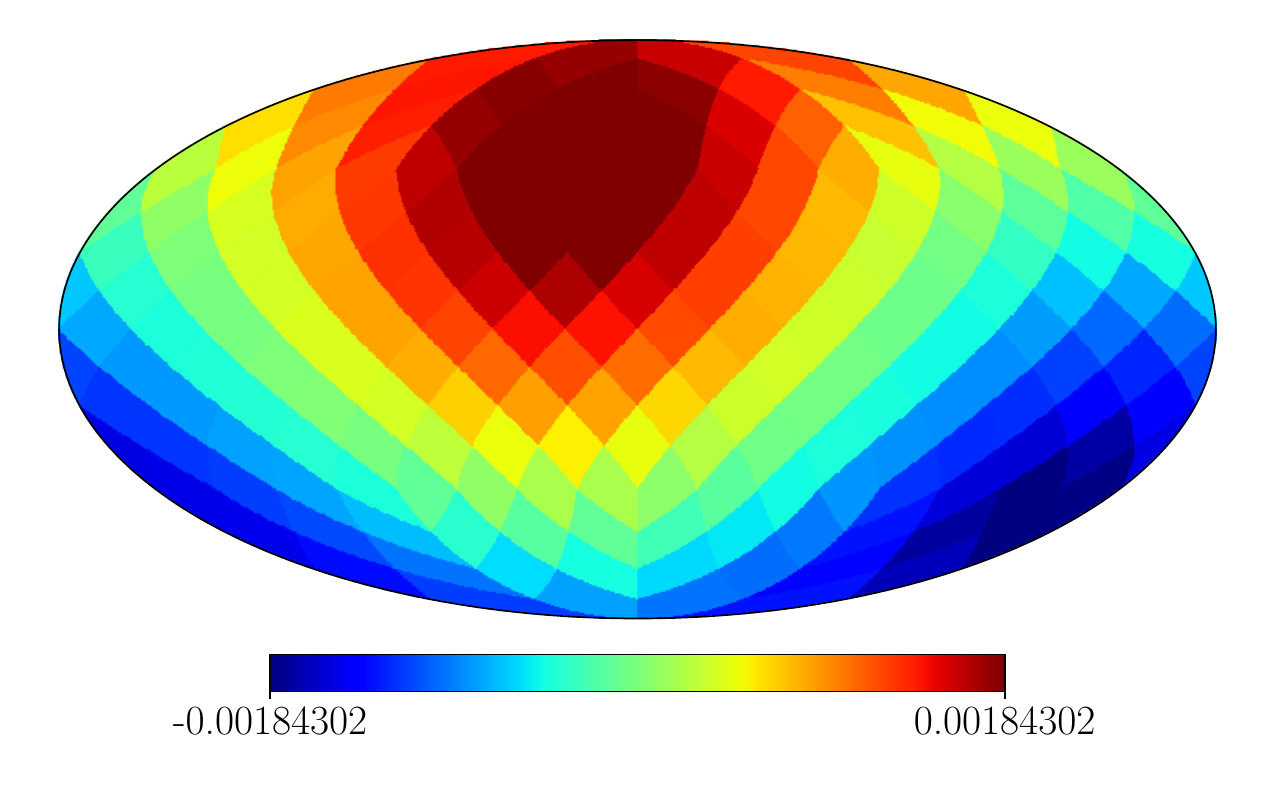}\includegraphics[width=0.35\textwidth] {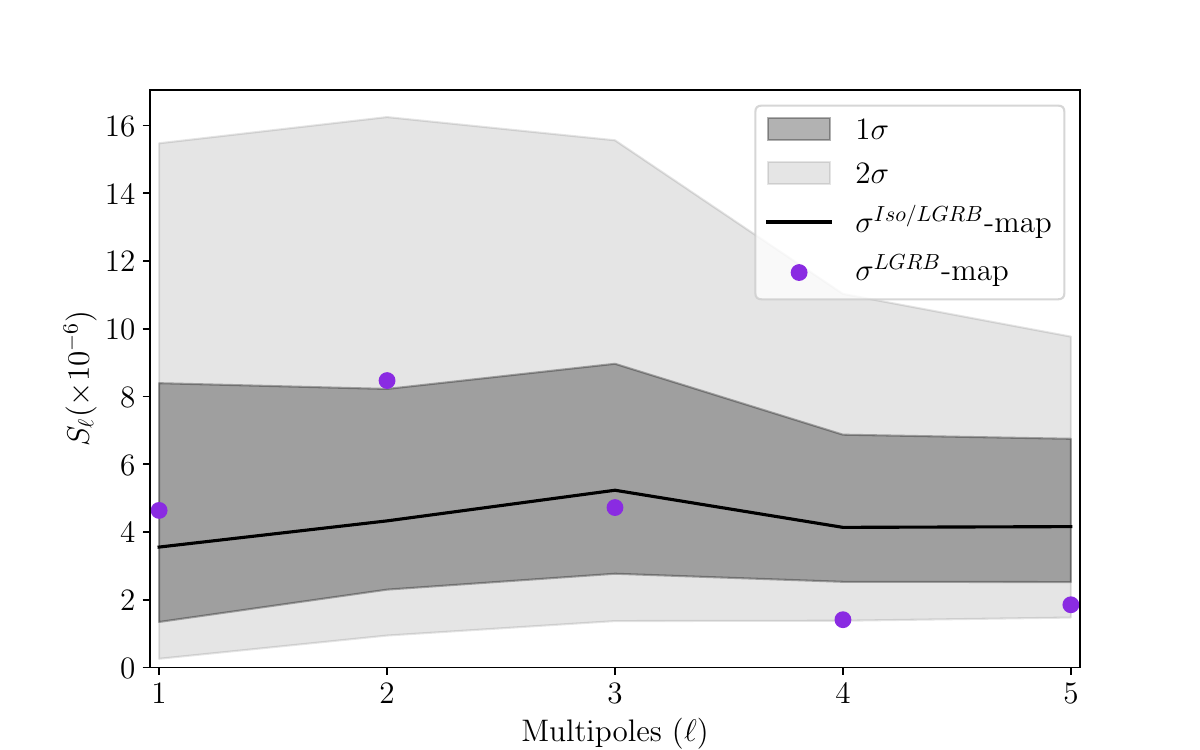}
\end{center}
\caption{Directional analysis of the GRBs angular correlations: SGRB and LGRB cases.
The first/second row corresponds to the SGRB/LGRB angular distribution analyses.
\textbf{Left panel:}  $\sigma^\text{SGRB/LGRB}$-map 
obtained from the directional analyses of the SGRB and LGRB samples. 
\textbf{Middle panel:} 
Corresponding dipole component of the $\sigma^\text{SGRB/LGRB}$-map. 
\textbf{Right panel:} 
Power spectrum ${\cal S}_\ell^{\text{SGRB/LGRB}}$, $\ell = 1,..., 5$ of the 
$\sigma^\text{SGRB/LGRB}$-map shown as dots, and the median power spectrum of 500 isotropic $\sigma$-maps are shown on the solid line. The shaded region corresponds to $1\,\sigma$ and $2\,\sigma$ deviations from the isotropic case. 
}
\label{sigma-SGRB-LGRB}
\end{figure*}

Our results are displayed in Figure~\ref{sigma-SGRB-LGRB}. From left to right in the first (second) row, we show the analyses for the SGRB (LGRB) sample, that is the $\sigma^\text{SGRB}$-map ($\sigma^\text{LGRB}$-map), its dipole, and its angular power spectrum ${\cal S}_\ell^{\text{SGRB}}$ (${\cal S}_\ell^{\text{LGRB}}$) for $\ell=1,...,5$.  
More precisely, the right panels in both rows of Figure~\ref{sigma-SGRB-LGRB} compare the data power spectra, ${\cal S}_\ell^{\text{SGRB}}$ and ${\cal S}_\ell^{\text{LGRB}}$ (dots), with the corresponding sets of power spectra obtained from synthetic $\sigma^\text{ISO}$-maps. 
These $\sigma$-maps were produced from two sets of $500$ simulated isotropic maps for each case, i.e., SGRB and LGRB. 
The continuous line represents the median power spectrum, and the shaded areas represent the 1$\sigma$ and 2$\sigma$ statistical confidence levels, respectively. 
The results are consistent with those found in Section~\ref{GRB-sky-dist}. 
For completeness, we also study the $f$-maps of the SGRB and LGRB samples, although the corresponding figures are not displayed. 
In both cases, the analyses show that these $f$-maps are also compatible with the SI hypothesis.

%%%%%%%%%%%%%%%%%%%%%%%%%%%%%%%%%%%%%%%%%%%%%%%%%%%%%%%%%%%%%%%%%%%%%%%%%%%%%%%%%%

\section{Consistency test: Spherical cap size}\label{cap-size}

\renewcommand{\thefigure}{B.\arabic{figure}}
\renewcommand{\tablename}{Table} 
\renewcommand{\thetable}{B.\arabic{table}}
\setcounter{table}{0}
\setcounter{figure}{0} 

In this Appendix we perform isotropy analyses of the $f$-maps considering different
spherical cap sizes, namely 
$\gamma = 45^{\circ},\, 60^{\circ},\, 65^{\circ},\, 70^{\circ}$ (see 
Table~\ref{table:cap-size}). These results support our previous conclusions regarding the isotropic distribution of the FERMIGRB catalog.

\begin{table}[ht]
\caption{Dipole directions for the FERMIGRB catalog}
\centering
\scalebox{1}{
\begin{tabular}{c c c}
\hline
\hline
$\gamma (^\circ )$ & $l \pm 11.2 \, (^\circ )$ & $b  \pm 11.2 \, (^\circ )$ \\
\hline
 $90$  & $268.02 $  & $-2.80$ \\

45 & $267.63 $  & ~~$4.86 $ \\
60  & $266.01$  & $-2.66 $ \\
65 & $266.85$ & $-2.38$ \\
70 & $265.99$ & $-4.36$  \\

\hline
\vspace{0.02cm}
\end{tabular}
} 
\begin{minipage}{\columnwidth}
\justify
\textbf{Notes.} Dipole directions for the $f^{\text{FERMI}}$-maps obtained from FERMIGRB catalog and different spherical cap sizes.
\end{minipage}
\label{table:cap-size}
\end{table}

%%%%%%%%%%%%%%%%%%%%%%%%%%%%%%%%%%%%%%%%%%%%%%%%%%%%%%%%%%%%%%%%%%%%%%%%%%%%%%%%%%

\section{Directional variation of the number of GRB 
and the \texorpdfstring{$f$}{f}-map signatures}\label{pearson}

\renewcommand{\thefigure}{C.\arabic{figure}}
\renewcommand{\tablename}{Table} 
\renewcommand{\thetable}{C.\arabic{table}}
\setcounter{table}{0}
\setcounter{figure}{0}

\begin{figure}[!ht]
\includegraphics[width=0.75\columnwidth] {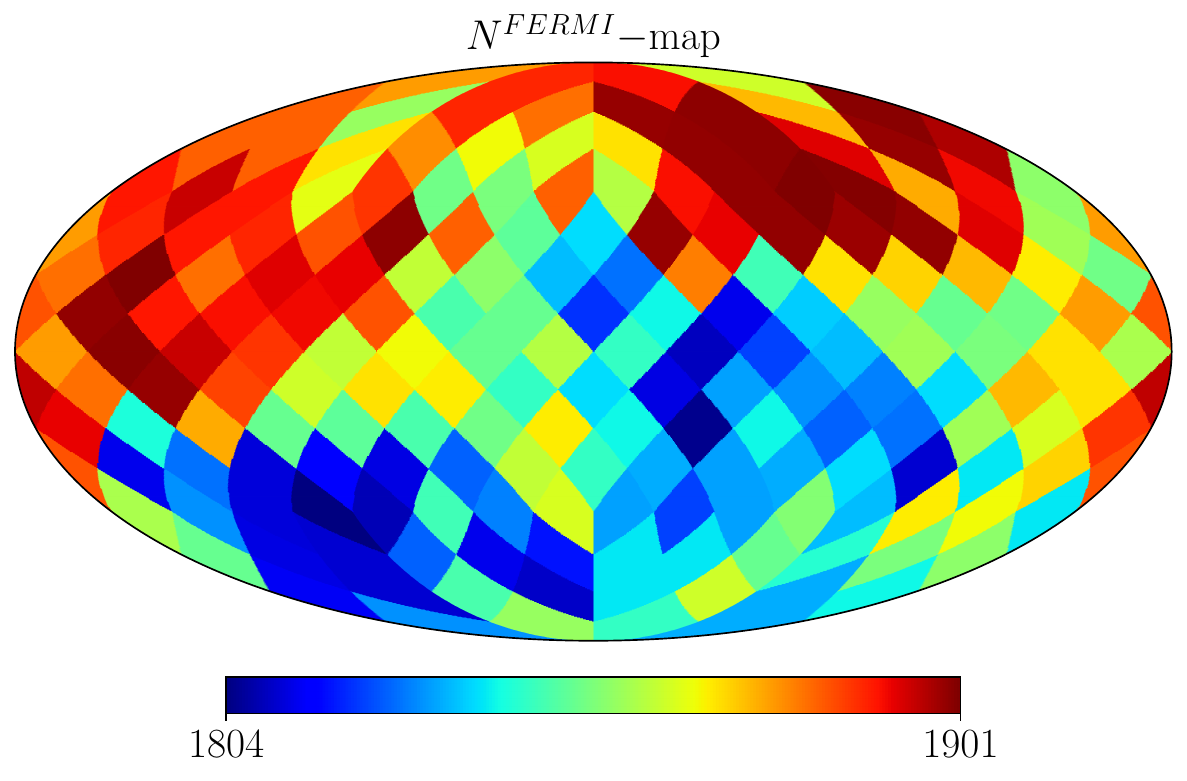}
\caption{Number-of-GRB-map from the  FERMI sample, or $N^{\text{FERMI}}$-map, shown in Mollweide projection and Galactic coordinates, to study the correlation analyses with the $f^{\text{FERMI}}$-maps (see Appendix~\ref{pearson} for details).}
\label{fig:number-GRBs}
\end{figure}

In this Appendix we investigate the possible correlation between the number of GRBs with the directional signatures of the $f$-maps. 
For this, we perform a correlation analysis of the number-of-GRB-map (that is, the map where the color in each pixel represents the number of GRB used to construct our $\sigma$-map or the $f$-maps, see Figure~\ref{fig:number-GRBs}) 
with the $\sigma$-map and with the $f$-maps, using the linear Pearson correlation coefficient $\cal{P}$. 
%    $|{\cal P}| = 0.198$
Our analyses are summarized in Table~\ref{table:Pearson}. 
According to the literature, for values of the Pearson coefficient in the interval (the vertical bars mean absolute value), 
$|\,{\cal P}\,| \in [0.0, 0.19]$ means that the linear correlation between 
the pair of maps is very low; for values $|\,{\cal P}\,| \in [0.6, 0.79]$ the correlation between the pair of maps is high; 
while for values $|\,{\cal P}\,| \in [0.8, 1.0]$ the correlation between the pair of maps is very high. According to these results, we conclude that the $N$-map, 
i.e., the number-of-GRB-map, that quantifies the number of GRB in each hemisphere, has a negligible correlation with the 
angular signatures found in the $f^{\text{FERMI}}$-map. 

\begin{table}[ht]
\caption{Pearson coefficients for the FERMI catalog and subsamples}
\centering
\scalebox{1}{
\begin{tabular}{l c c c}
\hline
\hline
 & $f^\text{FERMI}$-map & $f^{\text{FERMI$_{\text{MKD}}$}}$-map & $f^{\text{FERMI$_{\text{BAT}}$}}$-map  \\
\hline
$N$-map &  -0.076   &  -0.170  &  0.078  \\
$f^\text{FERMI}$-map & $-$ & 0.672 &  0.941 \\
\hline
\vspace{0.02cm}
\end{tabular}
} 
\begin{minipage}{\columnwidth}
\justify
\textbf{Notes.} Correlation analyses: Pearson coefficient calculated to study the linear correlation between different maps. For instance, the value $-0.170$ means that the $N$-map is 
weakly anti-correlated with the $f^{\text{FERMI}_{\text{MKD}}}$-map.
\end{minipage}
\label{table:Pearson}
\end{table} 

%%%%%%%%%%%%%%%%%%%%%%%%%%%%%%%%%%%%%%%%%%%%%%%%%%%%%%%%%%%%%%%%%%%%%%%%%%%%

\section{Statistical isotropy analyses of FERMIGRB fluence in BATSE channels}\label{fluence-fermi-batse}

\renewcommand{\thefigure}{D.\arabic{figure}}
\renewcommand{\tablename}{Table} 
\renewcommand{\thetable}{D.\arabic{table}}
\setcounter{table}{0}
\setcounter{figure}{0} 

\begin{figure*}[htbp]
\begin{center}
\includegraphics[width=0.32\textwidth]{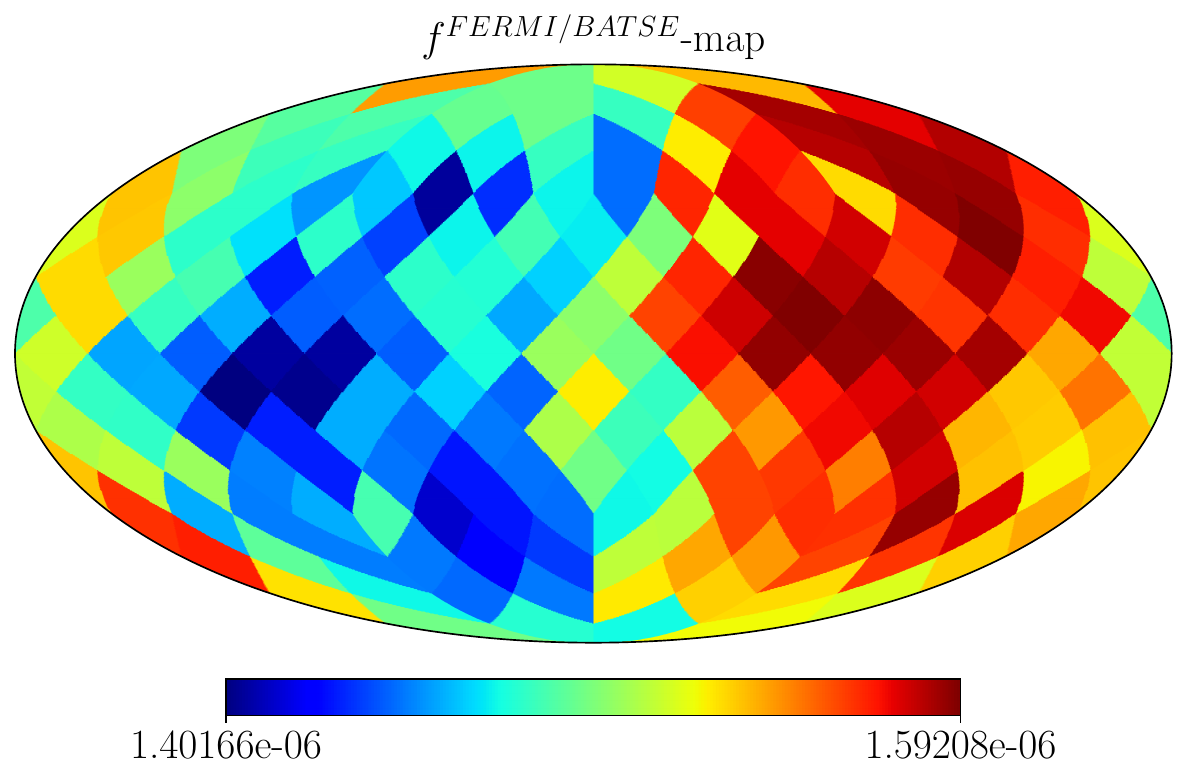}
\includegraphics[width=0.32\textwidth] {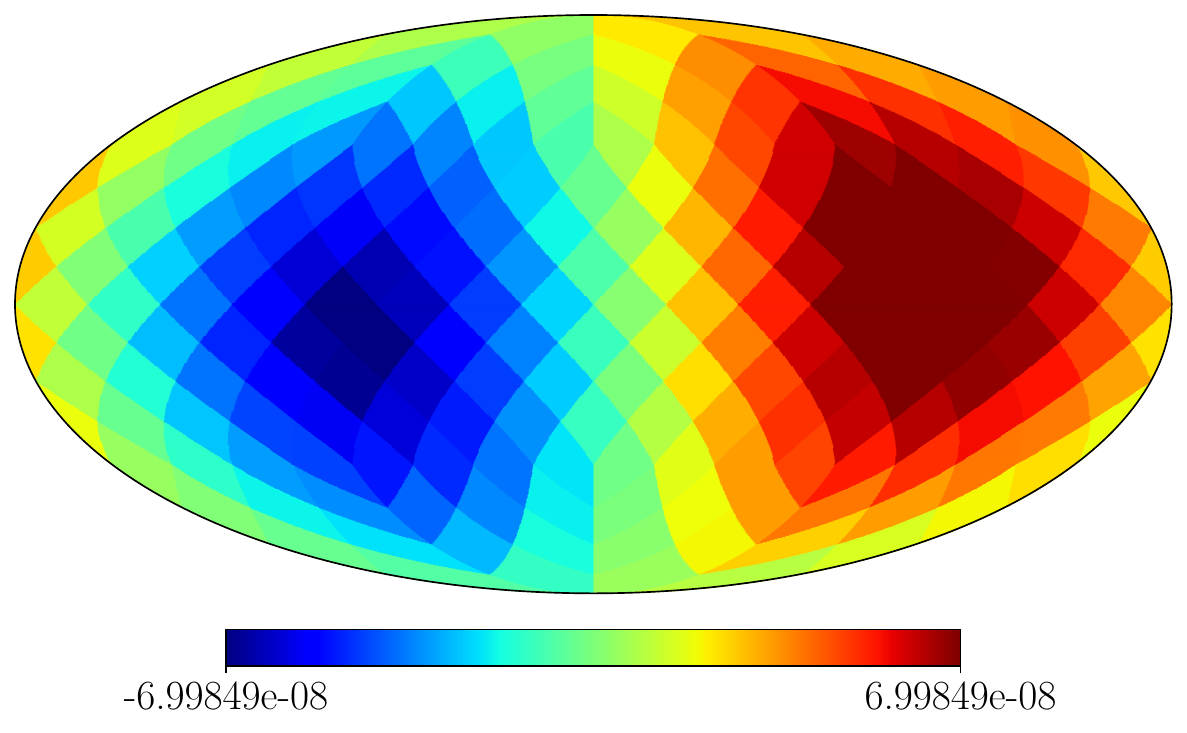}
\includegraphics[width=0.33\textwidth] {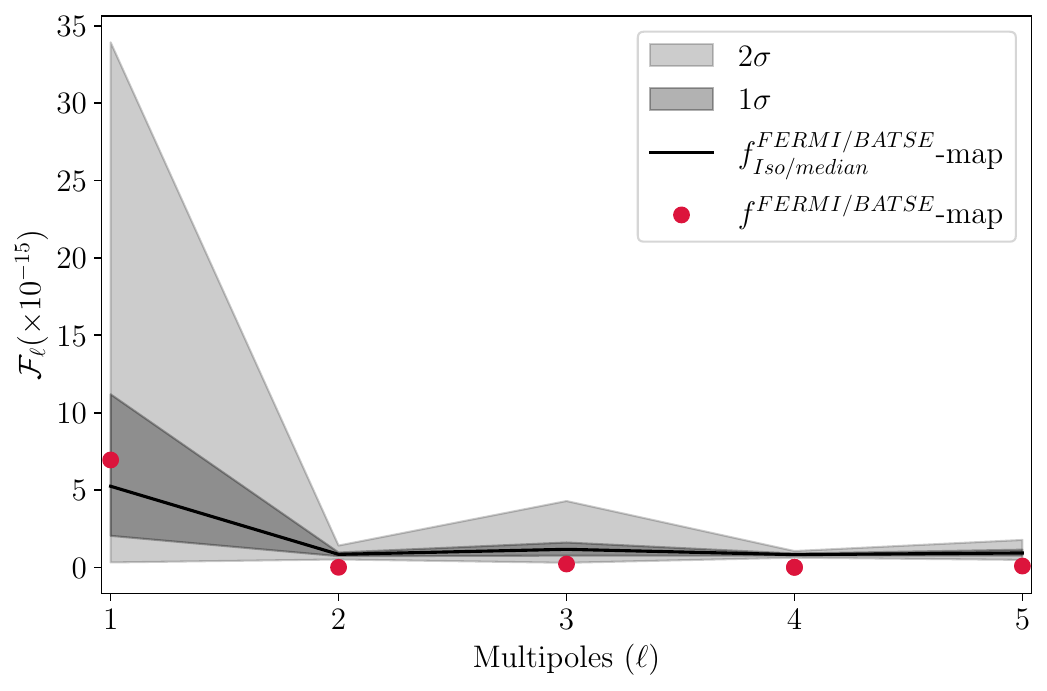} 

\includegraphics[width=0.32\textwidth]{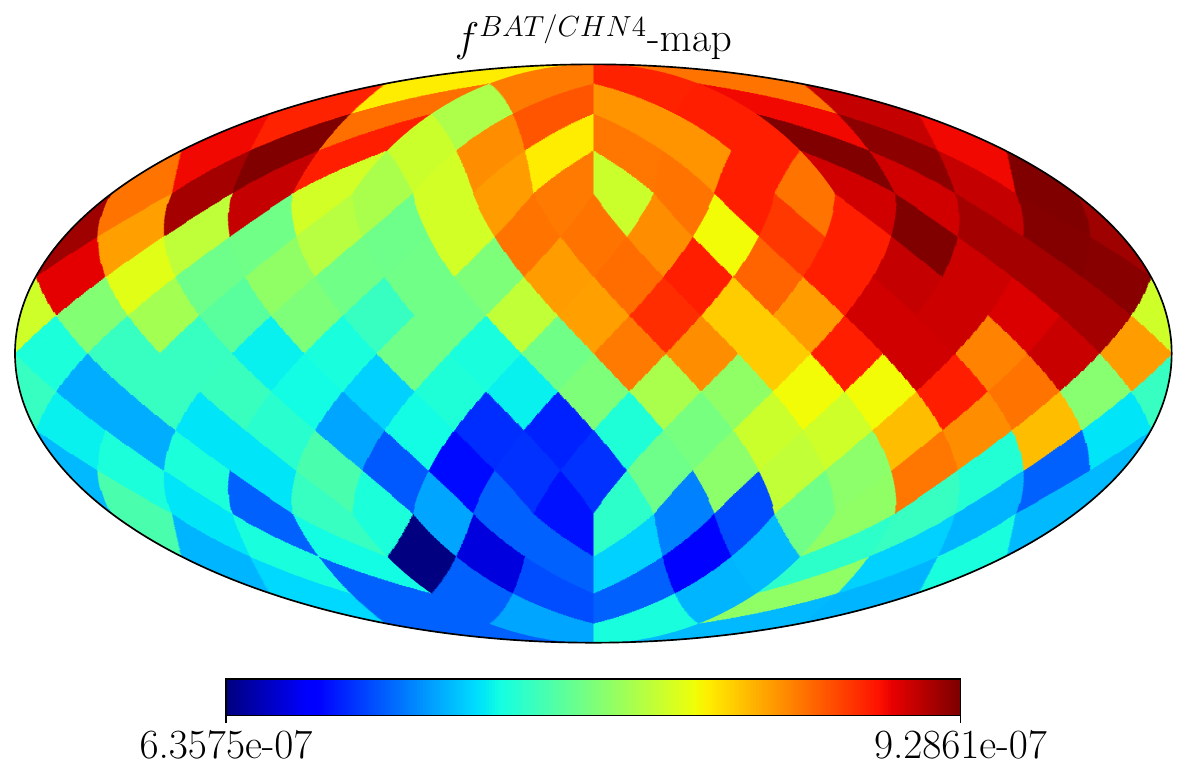}
\includegraphics[width=0.32\textwidth] {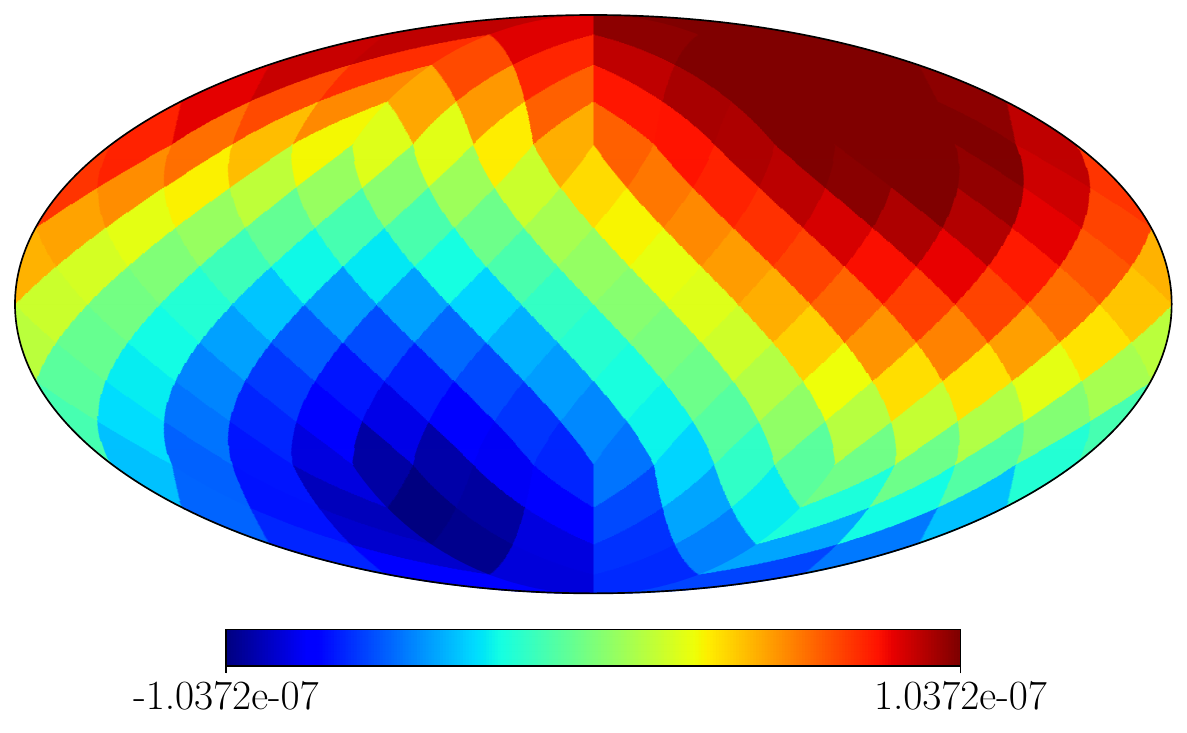}
\includegraphics[width=0.33\textwidth] {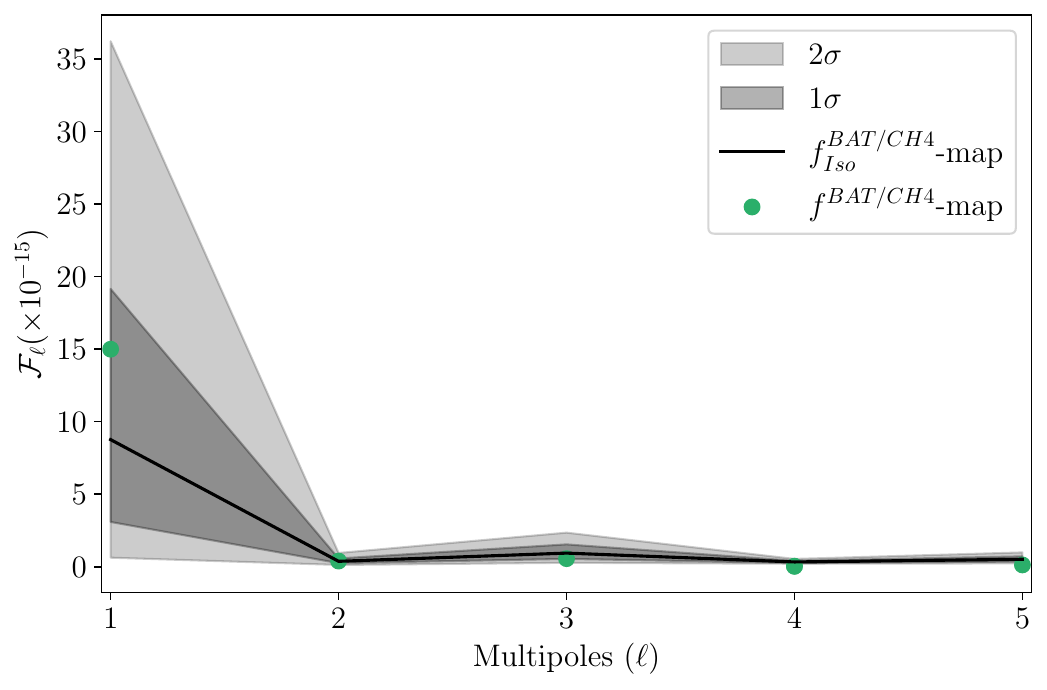}
 \end{center}
\caption{FERMI-BATSE analyses of the fluence angular distribution. 
{\bf Left panel:}   $f^{\text{FERMI/BATSE}}$-map 
for the FERMI-BATSE catalog. 
{\bf Middle panel:}
 Dipole component of the $f^{\text{FERMI$/$BATSE}}$-map. 
{\bf Right panel:} Power spectrum of the 
$f^{\text{FERMI$/$BATSE}}$-map. } 
\label{f-BATSE-maps}
\end{figure*}

Additionally, we study the statistical isotropy of the FERMIGRB fluence in the BATSE Channels. 
The power spectrum shown in the right panel of 
Figure~\ref{f-BATSE-maps} confirms the SI at $1\,\sigma$ confidence level, a fully similar result presented in Figure~\ref{figure:f_map_fermi} corresponding to the statistical analyses of the FERMIGRB catalog. 
Our results for the dipole directions of these analyses are summarized in Table~\ref{table:fluence-results}, and illustrated in Figure~\ref{fig:maps-mollw-results}.

\begin{figure*}[ht]
\begin{center}
\includegraphics[height=4.6cm,width=5.5cm]{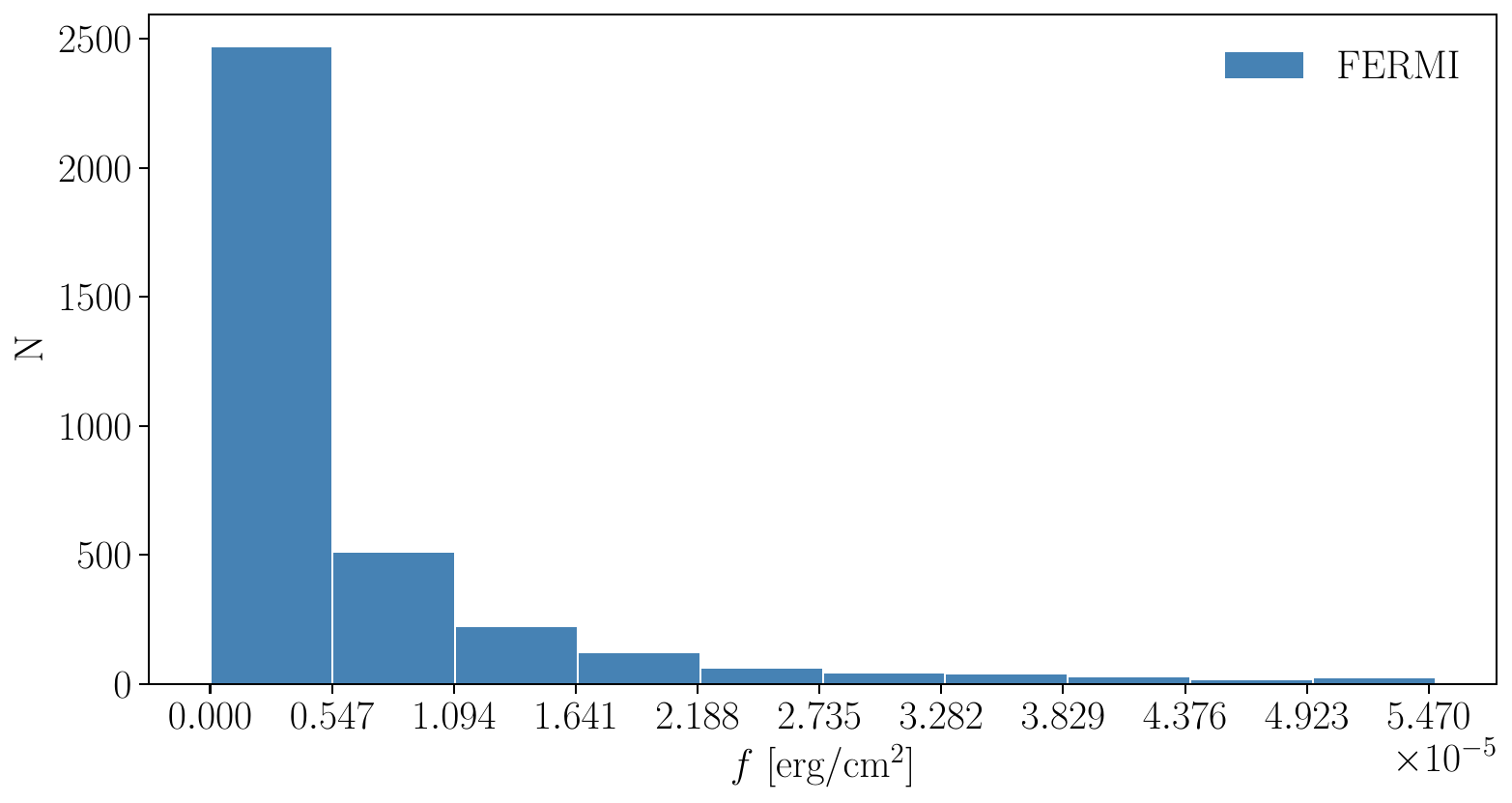}
\includegraphics[height=4.6cm,width=5.5cm]{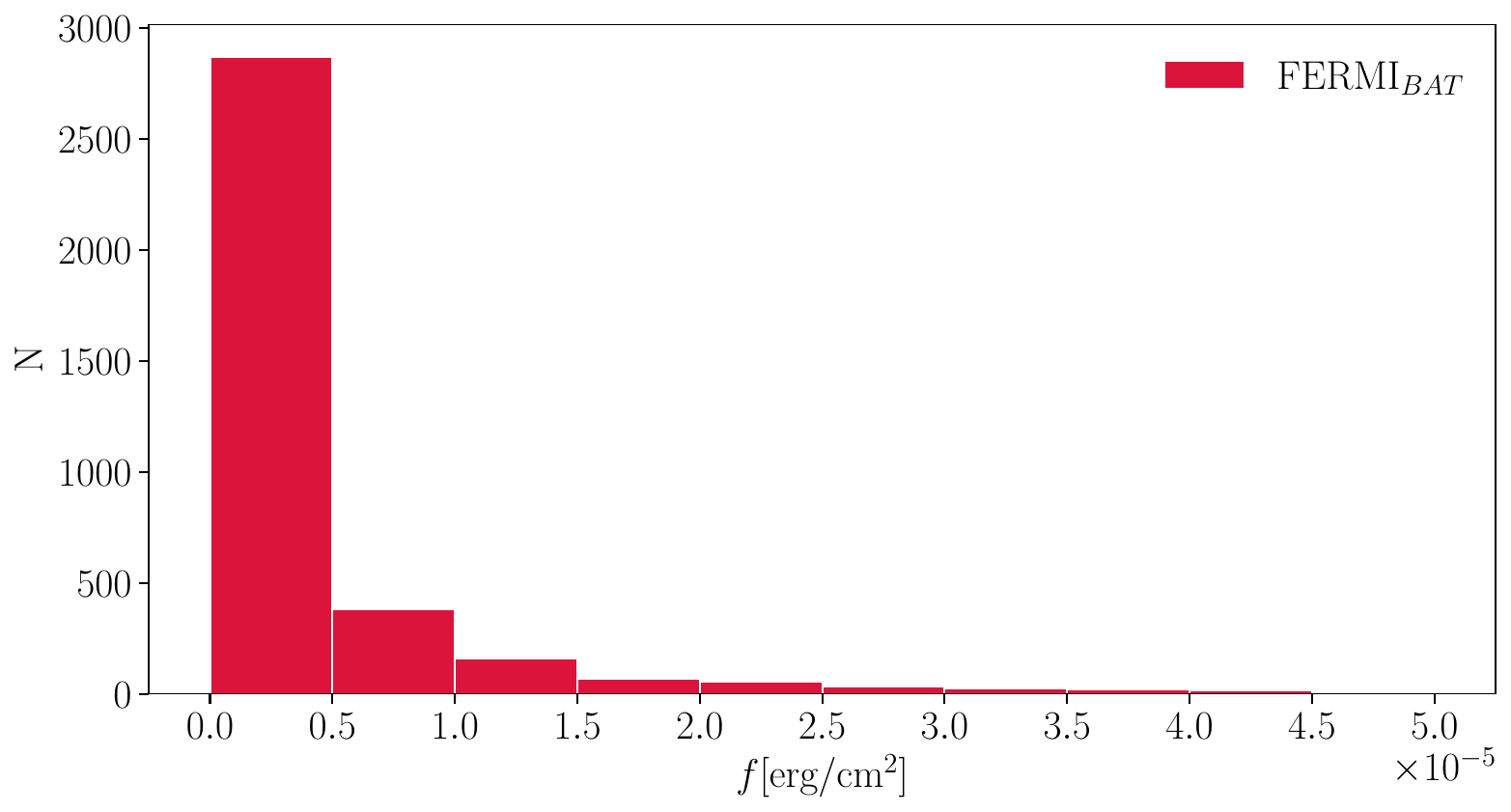}
\includegraphics[height=4.6cm,width=5.5cm]{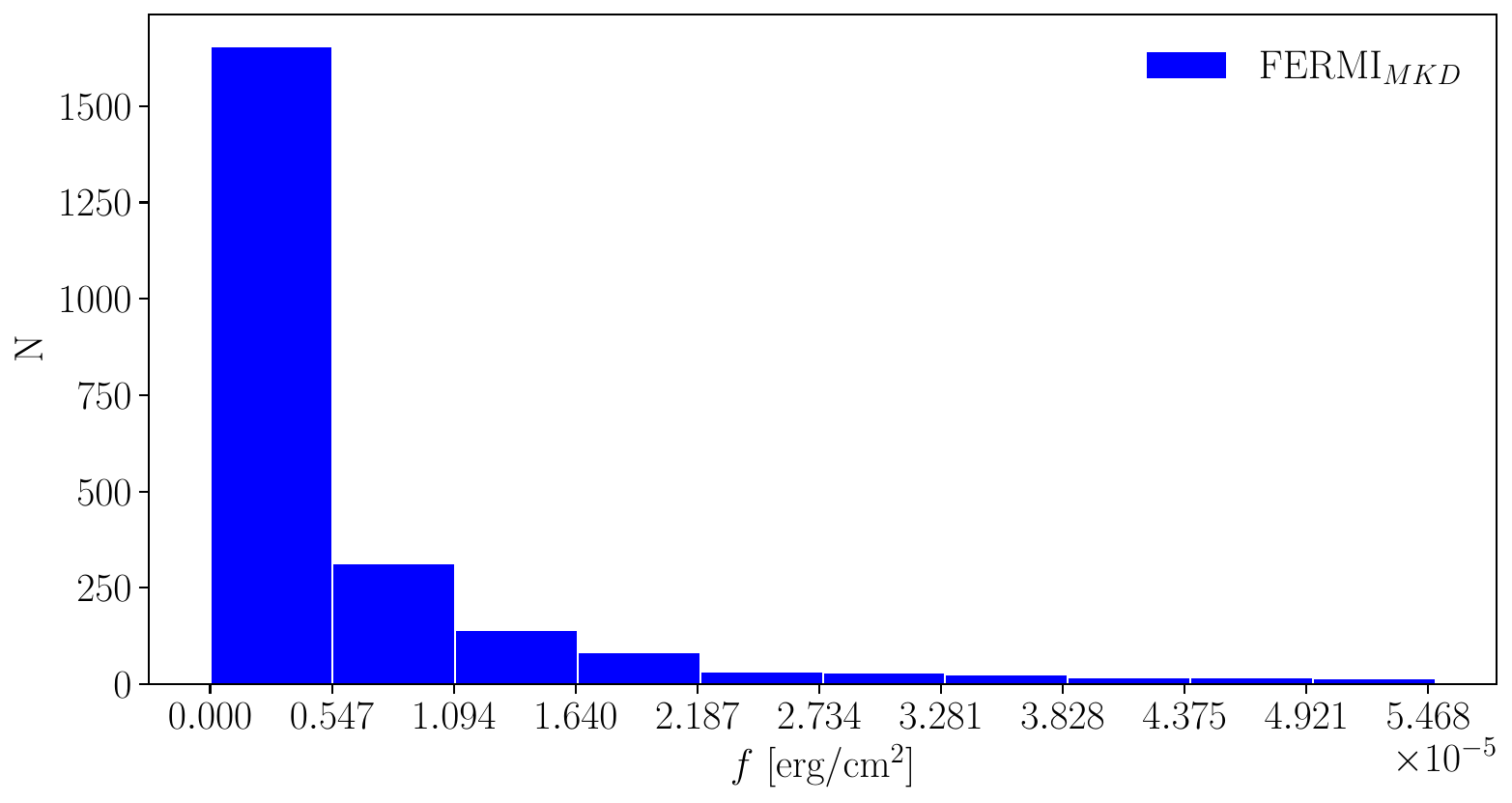}
\end{center}
\vspace{-0.2cm}
\caption{Fluence distributions of the FERMIGRB, FERMIGRB$_{\text{BAT}}$, and FERMIGRB$_{\text MKD}$ samples. \textbf{Left panel:} Frequency histogram of the measured fluence for the sample of $3691$ FERMIGRB, the bin size is $\simeq 3.02\times 10^{-5}$ [erg/cm$^{2}$], in the energy band, nominally 10-1000 keV. 
For visualization purposes, 12 GRB were removed from the original sample of 3703 GRB. 
\textbf{Middle panel:} Frequency histogram of the  FERMIGRB$_{\text{BAT}}$ sample in the BATSE standard energy band, nominally 50-300 keV. 
\textbf{Right panel:} Frequency histogram of the 2428 GRB remaining of the original FERMIGRB sample (left panel) after we remove the GRB located in the Zone of Avoidance, that is, those GRB with $|b| \le 20^{\circ}$ 
(i.e., $-20^{\circ} \le \text{DEC} \le +20^{\circ}$, see Figure~\ref{f-masked-maps}).
}
\label{fig:hist-fluence}
\end{figure*}

In Table~\ref{table:fluence-results} we present the direction, in Galactic coordinates, along which the fluence dipole appears.  In Figure~\ref{fig:maps-mollw-results} we illustrate all those directions in a Mollweide celestial sphere projection. We also show, in Figure~\ref{fig:hist-fluence}, the histograms of the fluence data for the catalogs analysed.

\begin{table}[!ht]
\caption{Dipole directions for the FERMI and BATSE catalogs}
\centering
\mbox{\hspace{-0.2cm}
\scalebox{1}{
\begin{tabular}{c c c} 
\hline
\hline
Sample  & $l \pm 11.2 \,\, (^\circ )$ & $b \pm 11.2 \,\, (^\circ )$  \\
\hline
FERMIGRB   & $268.02 $  & $\!-2.80 $    \\
FERMI$_{\text{MKD}}$ & $315.33$  & $5.73 $  \\
FERMI$_{\text{BAT}}$ & $254.41$ & $7.39$ \\

BAT/CHN1  & $167.96$ & $\!\!-16.19$ \\
BAT/CHN2 & $235.72$ & $\!\!-35.45 $ \\
BAT/CHN3 & $283.60$ & $10.69 $  \\
BAT/CHN4 & $249.99$ & $55.92 $ \\
BATSEGRB  & $240.46$ & $43.97$ \\
\hline
\vspace{0.02cm}
\end{tabular}
}
}
\begin{minipage}{\columnwidth}
\justify
\textbf{Notes.} Dipole directions for the $f$-maps obtained from different samples, with hemispheres $\gamma=90^\circ$. The locations of these directions on the celestial sphere are shown in Figure \ref{fig:maps-mollw-results}.
\end{minipage}
\label{table:fluence-results}
\end{table}

\end{appendix}

\end{document}